\documentclass[conference]{IEEEtran}
\ifCLASSINFOpdf
\else
\fi
\usepackage[english]{babel}
\usepackage[dvipsnames]{xcolor}
\usepackage[normalem]{ulem}
\usepackage{color,soul}
\usepackage{enumitem}
\usepackage{graphicx}
\usepackage{caption}
\usepackage{subcaption}
\usepackage[font={small}]{caption}
\usepackage{amsmath}
\usepackage{lipsum}
\usepackage{amssymb}
\usepackage{amsthm}
\usepackage{mathtools}
\usepackage{setspace}
\usepackage{ulem}
\usepackage{cancel}
\newcommand\numberthis{\addtocounter{equation}{1}\tag{\theequation}}
\usepackage{comment}
\usepackage[hidelinks]{hyperref}

\begin{document}
\bstctlcite{IEEEexample:BSTcontrol}

\title{Coverage Performance of UAV-powered Sensors for Energy-neutral Networks with Recharging Stations\\
	{\normalfont\large 
		\textbf{Oktay Cetinkaya}\IEEEauthorrefmark{1}\hspace{5mm}\textbf{Mustafa Ozger}\IEEEauthorrefmark{2}\hspace{5mm}\textbf{David De Roure}\IEEEauthorrefmark{1}%\textbf{D~E}\IEEEauthorrefmark{1}\IEEEauthorrefmark{2}\IEEEauthorrefmark{3}
	}\\%[-0.1ex]
}

\author{
		\IEEEauthorblockA{%
		\hspace{-0.3cm}  \IEEEauthorrefmark{1}Oxford e-Research Centre (OeRC)\\
		\hspace{-0.2cm} Department of Engineering Science\\
		\hspace{-0.2cm} University of Oxford, UK \\
		%\hspace{-0.2cm} \{engs2436, oerc0033\}@ox.ac.uk
  \hspace{-0.2cm} oktay.cetinkaya@eng.ox.ac.uk 
  %\hspace{-0.2cm} david.deroure@oerc.ox.ac.uk 
	}%\vspace{-10mm}
\and
	\IEEEauthorblockA{
		\hspace{-0.6cm}  \IEEEauthorrefmark{2}School of Electrical Engineering\\ \hspace{-0.5cm} and Computer Science\\
		\hspace{-0.6cm} KTH Royal Institute of Technology, Sweden \hspace{-0.3cm}\\
		\hspace{-0.3cm} ozger@kth.se\hspace{0cm}
	}
}

\maketitle

\begin{abstract}

%The projected number of Internet of Things (IoT) devices makes maintenance an issue since replacing/replenishing non-deterministically depleting innumerable batteries is not straightforward. Battery-less IoT is not a fantasy, yet the~devices should be somehow energized, either locally or remotely. Unmanned aerial vehicles (UAVs) can respond to this quest via wireless power transfer (WPT); however, to achieve energy-neutrality across the IoT networks and thus alleviate the maintenance issues, the sensor-destined UAV energy must be supplied by recharging stations having multi-source energy harvesting (EH) capability. Yet, this may cause the outage of sensors due to their sole dependence on UAV-delivered energy, which is absent when the UAVs have to visit recharging stations for battery replenishment. Hence, besides the UAV parameters (e.g., battery size and velocity), the duration of recharging and the distribution of stations providing it become crucial to minimize sensor outages. Regarding these metrics, this paper derives the coverage probability of UAV-powered sensors based on stochastic geometry. The analyses, elaborating on several trade-offs, reveal the fundamental design guidelines for energy-neutral IoT networks with recharging stations in regard to regulatory organization limitations, practical rectenna and UAV models, and the minimum power requirements of sensors.

The projected number of Internet of Things (IoT) sensors makes battery maintenance a challenging task. Although battery-less IoT is technologically viable, the sensors should be somehow energized, either locally or remotely.~Unmanned aerial vehicles (UAVs) can respond to this quest via~wireless power transfer (WPT). However, to achieve energy neutrality across the IoT networks and thus mitigate the maintenance issues, the UAVs providing energy and connectivity to IoT sensors must be supplied by recharging stations having multi-source energy harvesting (EH) capability. Yet, as these sensors rely solely on UAV-transferred power, the absence of UAVs causes sensor outages and hence loss of coverage when they visit recharging stations for battery replenishment. % Yet, this may cause the frequent outage of sensors~due to their sole reliance on , which is absent when the UAVs have to visit recharging stations for battery replenishment. 
Hence, besides the UAV parameters (e.g., battery size and velocity), recharging duration and station density %become crucial to minimize these outages. Considering these metrics, %this paper derives the coverage probability of UAV-powered sensors based on stochastic geometry. 
%Hence, UAV parameters such as battery size, velocity, recharging duration and~station density 
must be carefully determined to avoid these outages. %and provide coverage to sensors through UAVs. 
To address that, 
this paper uses stochastic geometry to derive the coverage probability of UAV-powered sensors. Our analysis sheds light on the fundamental trade-offs and design guidelines for energy-neutral IoT networks with recharging stations in regard~to the regulatory organization limitations, practical rectenna and UAV models, and the minimum power requirements of sensors.

%Therefore, minimizing sensor outages requires careful consideration of factors such as UAV parameters (e.g., battery size and velocity), recharging duration, and station density. To address this, this paper employs stochastic geometry to derive the coverage probability of UAV-powered sensors, including analyses of associated trade-offs. The results provide design guidelines for energy-neutral IoT networks with recharging stations, taking into account regulatory organization limitations, practical rectenna and UAV models, and minimum power requirements of sensors.

%\textcolor{red}{\lipsum[6-7]}
\end{abstract}


\section{Introduction}
\label{Sec:Intro}

Unmanned aerial vehicles (UAVs) are becoming increasingly essential %component of 
in wireless networks, %in which they act 
serving as either mobile users or base stations/access points (APs) \cite{baltaci2021survey}.
%Unmanned aerial vehicles (UAVs) are becoming an indispensable part of wireless networks as mobile users and base stations/access points (APs) \cite{baltaci2021survey}. 
In addition to the numerous value-added services they already provide, 
%Beyond the numerous value-added services introduced already, 
UAVs %are also instrumental in alleviating the battery constraints of the Internet of Things (IoT) via wireless power transfer (WPT)
also play a critical role in addressing the battery constraints of the Internet of Things (IoT) through wireless power transfer (WPT) \cite{cetinkaya2020efficient}. They are particularly beneficial for networks that rely on a multitude of sensors, which require excessive maintenance, or that operate in hard-to-reach areas, such as rainforests. By offering line-of-sight (LoS) air-to-ground links, UAVs enable highly efficient WPT \cite{yuan2021joint}, which, in turn, enables the battery-less operation of sensors deployed on the ground.
%The line-of-sight (LoS) air-to-ground link the UAVs offer is a key enabler for high-efficient WPT \cite{yuan2021joint}, which facilitates the battery-less operation of ground sensors. %deployed over ground 
%thanks to the %via 
After delivering the energy required by sensors, %for their task execution. %transmissions in various applications. % such as data collection from remote field. %increasing also the coverage probability. 
%More specifically, after the UAVs deliver power to the sensors, 
%Upon energy delivery, they 
UAVs can also collect sensory data %, i.e., they 
by acting as mobile APs, thereby offering an all-in-one solution. In these settings, UAVs govern both energy and data flows with no 
human supervision for battery maintenance or terrestrial APs for data collection, aiming for a certain level of autonomy in network operation. 

Since the primary goal of a WPT-enabled UAV setting is to minimize the battery %and the associated maintenance 
constraints of sensors, besides the prevailing energy scarcity across the IoT domain, %the source that UAVs 
the energy required for UAV operation, including the power transferred to sensors, has to be provisioned %to realize \textit{energy-neutrality} of the system.
within the network. In this way, \textit{energy neutrality} can be enabled %across the IoT domain 
\cite{long2018energy}, mitigating the challenges mentioned above. %, such as grid dependency and limited battery lifetime. %“generated" %since the opposite case would be similar to driving an electric vehicle run by a petrol generator. Hence, the UAVs require 
%The ultimate goal of such systems is to minimize the gap between the harvested energy and the consumed energy within the network, especially across the IoT domain \cite{long2018energy}. 
%Energy neutrality along with WPT mitigates everlasting constraints such as batteries. 
One approach for achieving this goal is to replenish UAV batteries via recharging stations having energy harvesting (EH) capability. Here, multiple ambient sources, such as solar and wind power, can be simultaneously exploited \cite{HEH} to minimize the variance and intermittency in the EH output for assured reliability. %Such architecture is a promising candidate to fulfill the introduced energy-neutrality vision for the UAV-powered sensors \cite{cetinkaya2020internet}.

%There are countless examples in the 
The literature has vast examples of UAV-based service~provisioning for IoT devices, including recharging stations. In most cases, the UAVs operate as flying APs \cite{alzenad20173} to collect sensory data underpinned by %providing coverage to sensors, underpinned by 
terrestrial counterparts. They sometimes deliver power \cite{xie2021uav} in addition to or aside from AP functionality \cite{ye2020optimization}. The recharging stations are usually deemed to have mains connection, or their source of power is untold \cite{qin2020performance}, both referring to a case with an unlimited energy source, i.e., the total disregard of the energy neutrality objective.  

One domain that has been exhaustively studied in the literature is the UAV-assisted IoT networks, in which the~UAVs provide coverage to sensors. For example, the authors in~\cite{zhang2022energy} optimized the trajectory of a UAV energized by a solar-powered recharging station in consideration of data rate, energy consumption, and fairness of coverage. % in the UAV-assisted IoT networks, 
%where the UAV is energized by a solar-powered recharging station. 
However, they focused on a single UAV operation without considering the effect of recharging station density, limiting their application potential. Furthermore, the authors in \cite{hassija2020adistributed} proposed a distributed blockchain-based scheme to enable secure and reliable energy exchange between UAVs and recharging stations. %which could realize UAV-assisted IoT applications. 
However, they did not consider the energy-neutral operation of their system, leading to an impractical solution. %In \cite{reddy2021efficient}, UAV-assisted IoT networks were investigated via energy-efficient management schemes, such as deep reinforcement learning-based channel and power allocation algorithms. 
The authors in \cite{chu2022joint} proposed reinforcement learning algorithms to jointly optimize the velocity and energy replenishment of UAVs that collect data from sensors. Although they enabled efficient transfer learning techniques to decrease the learning time and improve the overall learning process, they did not take energy neutrality into consideration in their setting, similar to other studies.
%The main contribution of this work was enabling efficient transfer learning techniques to decrease the learning time and improve the overall learning process. However, they did not take into consideration of energy-neutrality in their setting, similar to other studies. %the energy-neutral operation of the UAV-assisted IoT networks. 

As discussed in \cite{wei2022uav} and the references therein, the literature on UAV-assisted sensor coverage mainly focused on various other aspects, such as clustering the sensor nodes for more energy-efficient data collection, different flying modes of UAVs, and joint path planning and resource allocation via graph-theory, optimization, machine learning, etc. Despite the promising findings of these studies, the research must look towards a more pressing and fundamental issue, i.e., how to achieve energy-neutral IoT at the minimal sensor outage, occurring due to recharging station-driven UAV operation. %not the only aspect that has been disregarded by the existing literature, despite the immense impact it has on the operation. %, which inevitably causes a discrepancy between the theory and practice. 
%To be specific,
%considerations need to be adopted not only for minimizing the discrepancy between the theory and practice created by the existing studies but also for achieving energy-neutrality in the IoT domain: 
Hence, the following aspects need considerable attention for a more realistic analysis of the coverage performance of sensors, facilitating energy neutrality in the IoT domain: 
\textit{i)} effective isotropic radiated power (EIRP) limitations enforced by regulatory organizations; \textit{ii)} practical rectenna models with non-linear EH behavior; \textit{iii)} minimum power requirements of sensors; \textit{iv)} individual duration of each UAV operation; and most importantly, \textit{v)} a limited source of power for WPT, i.e., the UAVs energized by multi-source EH recharging stations.
%These help us to narrow down the focus of this study to the current gap in the field: an autonomous IoT network driven by UAVs with WPT and AP capability, the operation of which is powered by multi-source EH recharging stations, cutting the deficit between the generated and consumed energy across the network. 

%The current gap in the research field being highlighted, we take a step forward from our previous study \cite{cetinkaya2020efficient} to investigate the coverage probability in such settings by considering the aspects explained above. Our model consists of energy-neutral UAVs delivering power and collecting data. Since no terrestrial APs considered, no service available during the trip and battery replenishment, which refers to outage or no coverage. To goal is to maximize coverage probability by tweaking recharging station density, battery capacity.

%The current gap in the research field being highlighted, this work aims to convey a high-level discussion on the coverage probability of sensors powered by multi-source EH recharging stations through directive UAVs. Since the UAVs singlehandedly manage both energy and data flows in our envisioned scenario, no service is  available during their trip and battery replenishment, which refers to outage \textit{or} no coverage. The goal is to maximize the probability of coverage by tweaking the crucial design parameters, such as the station density, UAV velocity, battery capacity, and recharging time. 

The current gap in the research field 
being highlighted, this work aims to %convey a high-level discussion by 
mathematically analyze the coverage probability of sensors powered by multi-source EH recharging stations through UAVs performing WPT with directive antennae. Since the UAVs manage both energy and data flows in the envisioned scenario, no service is available during their trip (towards sensors and recharging stations) and battery replenishment, which refers to an outage \textit{or} no coverage of sensors. The goal is to maximize the coverage probability by tweaking the crucial design parameters, such as the station density, UAV velocity, battery capacity, and recharging time. 

Following the agenda given above, we used stochastic geometry in order to derive a tractable expression for the event of service guaranteeing a certain level of coverage. 
%we used stochastic geometry and derived a tracktable expression for the service, and hence the coverage probability. 
Our analyses, based on practical rectenna and UAV propulsion models, revealed the non-trivial relationships between the UAV attributes (e.g., availability, velocity, descent altitude, antenna directivity, energy budget, output power, transmission duration, and operating frequency); sensor characteristics (e.g., sensitivity, antenna gain, and power conversion efficiency);  medium specifications (e.g., urban, suburban); and application requirements (e.g., minimum reporting frequency). 
In the end, our study provides an upper bound for the coverage probability of sensors, which can be practically achieved by carefully selecting the design parameters in a UAV-powered energy-neutral application scenario involving EH recharging stations.

%In summary, we are doing something similar to \cite{qin2020performance}, but 
%Novelty/contributions:
%\begin{itemize}
    %\item energy-constrained/limited charging stations (EH can be defined as a non-deterministic process)
    %\item UAV movement on the horizontal plane, $h_l$ (more realistic)
    %\item non-linear RF EH eq. extracted from empirical measurements (more realistic)
    %\item FCC regulations, reporting frequency requirements (more realistic)
    %\item No TBS. Completely UAV-driven, energy-neutral IoD scenario, where the UAV(s) provide also service
    %\item Unique investigation/analysis of optimum altitude, required min. outage, req. reporting frequency, availability, area, number or sensors to be supported (or their density), and so on.
%\end{itemize}

%\newpage
The remainder of this paper is organized as follows. We first introduce the system model in Section~\ref{Sec:SystemModel}, where the event of service, incorporating directivity, UAV propulsion, WPT and rectenna models, and FCC regulations, are formulated. %and the fundamental assumptions are explained. %Accordingly, the (maximum) sensing range of wireless devices is derived. Sec.~\ref{sec:Coverage} outlines the conventional deployment models and formulates an efficient deployment strategy based on CPP, providing a lower-bound for the required number of wireless devices (by using the sensing range devised). 
Then, in Section~\ref{sec:PA}, we derive the coverage probability of sensors based on stochastic geometry. This is followed by the numerical evaluation of the proposed model in Section~\ref{sec:results} to reveal the design guidelines that must be followed for the best achievable coverage while meeting application requirements. Finally, Section~\ref{sec:conclusions} discusses future research directions and concludes the paper.

%\textcolor{red}{\lipsum[5]}

%\vspace*{\fill}

%\textcolor{red}{\lipsum[4-9]}

\newpage
\section{System Model}
\label{Sec:SystemModel}

%  \item Stochastic geometry.

\begin{figure}[t!]
    \centering
    \includegraphics[width=\linewidth]{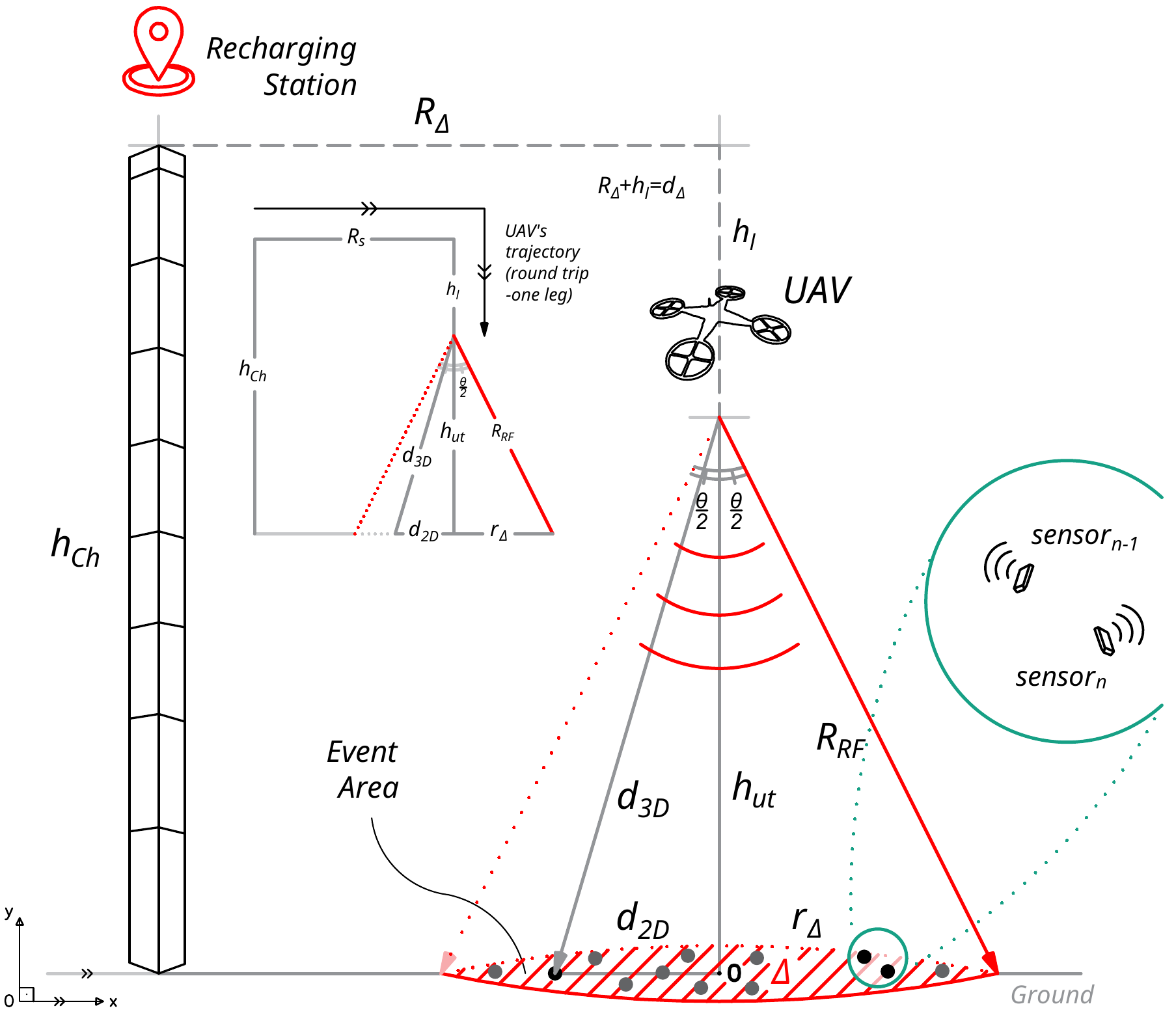}
    \caption{Illustration of the envisioned network scenario.}
    \label{fig:1}
    %\vspace{-5mm}
\end{figure}

As illustrated also in Fig.~\ref{fig:1}, we envision an energy-neutral network scenario, which comprises: \textit{i)} recharging stations, \textit{ii)} UAVs, and \textit{iii)} battery-less sensors. The UAVs retrieve energy from multi-source EH recharging stations via inductive power transfer, fly towards the sensors, energize them via radio frequency (RF) power transfer, and collect their data. During power transfer and data collection, which refers to \textit{service}, the UAVs do not move in the 3D space; they just hover at the centers of event areas, i.e., where sensors reside. The sensors become active as soon as they intercept enough power from a UAV. When active, they probe their vicinity for an application-defined parameter, e.g., temperature, humidity, and/or noise level, and notify their respective UAV with their readings. After collecting sensor data, the UAVs fly back to the nearest recharging station to replenish their batteries. 

As explained, the UAVs singlehandedly manage energy and data flows within the network in an autonomous manner; there are no other authorized entities, terrestrial APs, energy providers, etc. The service is provided at the centers of event areas, i.e., randomly located circles with %\xout{\textcolor{red}{fixed}} 
radius $r_\Delta$, which are modeled as a Poisson point process (PPP). %The centers of circles/event areas %, above which the UAVs are providing service, are modeled as a Poisson point process (PPP). 
Within each circle, the sensors are uniformly distributed, and finally, the locations of recharging stations are modeled as a PPP $\Phi_{Ch}$ with density $\lambda_{Ch}$. Below, we explain the service in detail.

%We consider a UAV-enabled cellular network composed of TBSs and UAVs, where the UAVs are located at the centers of hotspots. In order to model the locations of the users in the hotspot, one of the most popular models in literature is Poisson cluster process (PCP) [10]. There are two possible types of PCP: (i) Thomas cluster process and (ii) Matern cluster process (MCP). In this letter, we model the locations of the users in hotspots using MCP. In particular, the hotspots are modeled as randomly located disks with fixed radius rc. The centers of the disks, above which the UAVs are deployed, are modeled as a Poisson point process (PPP). Within each disk, the users are uniformly distributed. The UAVs are assumed to hover at a fixed altitude h above each hotspot center. The locations of the TBSs are modeled as a PPP TBS with density t. 

%Unlike existing literature, the main objective of this letter is to study the impact of the spatial distribution of the charging stations on the performance of the above setup. We model the locations of charging stations as a PPP c with density c.

\subsection{Service Provisioning}
\label{section:availability}
In our envisioned scenario, recharging stations refill the UAV battery with energy that is just enough %should spare enough energy to make it to the nearest charging station before they deplete their battery. 
for \textit{i)} a round trip to the event area, i.e., travel to its center and descent/ascent to/from it, \textit{ii)} providing \textit{service}, %i.e., power transfer to sensors in the event area and collection of their data, 
and \textit{iii)} hovering when providing service. During the trip and getting its battery replenished, the UAV cannot provide any service, i.e., it is unavailable. %, which means that no power transfer to the sensors in the event area, $\Delta$, and collection of sensor data.

%\begin{definition}[Availability probability]
%\textcolor{red}{We define the event A that indicates the availability of the UAV. 
The definition above confirms that the service is conditioned on the distance between the point where it is provided and the nearest charging station, $d_{\Delta} (=\!R_\Delta+h_l)$. However, only one of $d_\Delta$'s components, namely $R_\Delta$, changes randomly due to the distribution of recharging stations. Hence, the probability of UAV's availability, i.e., the event of service $\mathcal{E}$, is conditioned on $R_\Delta$, which can be given as: %(service is given as an event, $\mathcal{S}$):
\begin{equation} 
\label{eq:1}
\begin{split}
    P_{(\mathrm{e|R_\Delta})} & = \mathbb{P}(\mathcal{E}|R_\Delta)\\
    & = \frac{t_{PT}+t_{AP}}{t_{PT}+t_{AP}+t_{Ch}+t_J},
\end{split}
\end{equation}

\noindent where $t_{PT}$ is the time spent for power transfer, $t_{AP}$ is the time spent for data collection, $t_{Ch}$ is the time spent for recharging the UAV battery, and $t_J$ is the time spent for the round trip. %to travel to and from the nearest charging station.
%\end{definition}
Here, each of $t_J$, $t_{PT}$, and $t_{AP}$ can be defined as:
\begin{gather}
\label{eq:2}
\begin{aligned} 
t_J &=  \frac{2(R_\Delta + h_l)}{V}=\frac{2d_{\Delta}}{V}, \\
t_{PT} &= \frac{E_{PT}}{P_T},\\
t_{AP} &=  \frac{B_{\mathrm{UAV}}-t_{PT}(P_h+P_T)-t_JP_J}{P_h},
\end{aligned}
\end{gather}

\noindent where $P_J$~is the power consumption during the trip, $V$ is the UAV’s velocity during the trip, $E_{PT}$ is the energy budget spared for power transfer, $P_h$ is the power consumption during hovering at the center of an event area, $P_T$ is the transmit power of the UAV, and $B_{\mathrm{UAV}}$ is the energy level of the UAV battery. %which includes both the propulsion power and the total communication power. 
%Note that for power consumption during traveling, we focus on the power consumed to travel the horizontal and vertical distance $Rs$ and $h_l$ since they are typically larger than the power consumed during landing, after reaching the charging station.}\\
We should note that the UAV battery might not be fully charged always since it predominately depends on the power transfer rate of the wireless pad, $\xi_{Ch}$, of the recharging station and the time the UAV spends on it, $t_{Ch}$. %y also considering the sensitivity of rectenna, i.e. the minimum (threshold) power $P_{th}$ required for its activation,
From (\ref{eq:1}), we also know that $t_{Ch}$ is inversely proportional to UAV availability since it is on the denominator of the equation, so it should be limited. That is also because the UAV battery has a limited/maximum capacity, $B_{\mathrm{max}}$, so the UAV should not reside on the charging pad beyond when the battery gets fully charged, referring to $t_{Ch}^{sat}$. Note that the saturation time $t_{Ch}^{sat}$ can be achieved sooner or later depending on the charging rate $\xi_{Ch}$ since $B_{\mathrm{max}}$ is fixed. Considering all these, an accurate battery charging model for the UAV can be given as: %expressed as the following piecewise linear function:
\begin{equation}
	B_{\mathrm{UAV}}(t_{Ch}) = \begin{cases}
		\xi_{Ch} \cdot t_{Ch},&~t_{Ch} \in [0, t_{Ch}^{sat}],\\
		B_{\mathrm{max}},&t_{Ch} \geq t_{Ch}^{sat}.
	\end{cases}
\label{eq:battery_charging}
\end{equation}
%which is non-decreasing and continuous for all $B_{sta}\!\in\!\mathbb{R}$, $B_{sta}\geq0$.

\noindent and finally, by taking the expectation of %the conditional probability given in
(\ref{eq:1}), the service probability of the UAV can be calculated as:
\begin{equation}
P_\mathrm{e} = \mathbb{E}_{\Phi_{Ch}} \bigg [ \frac{t_{PT}+t_{AP}}{t_{PT}+t_{AP}+t_{Ch}+t_J} \bigg ].    
\end{equation}
%\textcolor{red}{These XXX}

%\begin{comment}
\subsection{Power Transfer}

The UAV performs RF power transfer with a directional antenna having a pencil-beam-like radiation pattern. For such an antenna, i.e. with one major lobe and very negligible minor lobes of the beam, the gain $G_T$ can be approximated by:
\begin{equation}
G_T =
\begin{cases}
\approx\!\dfrac{30000}{\theta_B^2},&~\frac{-\theta_B}{2} \leq \varphi \leq \frac{\theta_B}{2},~~~~~\text{(major lobe)}\\
~~~~~g(\varphi),              &~~~~~~~\text{otherwise},~~~\text{(minor lobes)}
\end{cases}
\label{eq:directivity}
\end{equation}
where $\varphi$ is the sector angle, $\theta_B$ is the directional antenna half-power beamwidth (HPBW) -both in degrees, $\approx\!30000/\theta_B^2$ is the maximum gain, and $g(\varphi)$ is the gain outside of the major lobe (including minor lobes), which can be neglected~\cite{balanis2015antenna}. Note that (\ref{eq:directivity}) is for a symmetrical radiation pattern, where the HPBWs in each plane are equal to each other, i.e. $\theta_{1d}\!=\!\theta_{2d}$.

Contrary to expectations, the transmit power of the UAV, $P_T$ in (\ref{eq:2}), cannot be altered casually; it is determined by
regulatory organizations, e.g., the Office of Communications (Ofcom) in the UK, the Federal Communications Commission (FCC) in the US. For example, FCC Part 15.247 rules \cite{2010} %\footnote{Federal Communications Commission CFR, Title 47, Vol. 1, Part 15, 2010. [Online]: https://www.govinfo.gov/app/details/CFR-2010-title47-vol1.} 
declare that the maximum $P_T$ fed into the (in our case, UAV's) antenna cannot exceed $30$dBm ($1$W) for the industrial, scientific, and medical (ISM) bands, in which the maximum effective isotropic radiated power ($\mathrm{EIRP_{max}}$) is limited to $36$dBm ($4$W). This indicates that increasing $G_T$ necessitates a proportional decrease in $P_T$, and vice versa, such that the total RF power radiated by the antenna remains the same, i.e., $4$W EIRP, where $\max(P_T)\!=\!1$W for each case. For directional dispersion, however, there are some exceptions to the $\mathrm{EIRP_{max}}$, details of which can be found in \cite{2010}. 

%\textcolor{blue}{For example, in the $2.4$GHz band, increasing $G_T$ to get an EIRP above $36$dBm is allowed (up to $52$dBm), where $P_T$ must be reduced by $1$dBm for every $3$dBi increase of $G_T$.}
%Thus, at $2.4$GHz, a power source increasing its EIRP (with $G_T$) has to decrease its $P_T$ to comply with the FCC regulations. As the UAV in our case has a fixed power transfer budget $E_{PT}$, decreasing $P_T$ allows a longer duration of power transmission $t_{PT}$, i.e. lengthened coverage lifetime. Although this sounds attractive, the duty cycle of the w-p\textit{D}s will accordingly be altered, which cannot be tolerated always due to the certain reporting frequency requirements of the IoT applications \cite{cetinkaya2017}. %\textcolor{blue}{Furthermore, since decreasing $P_T$ (increasing $f$) will accordingly lower $P_R$, the w-p\textit{D}s may need to switch from power-neutral to energy-neutral operation \cite{sliper2020}, which is not desired for this particular scenario.} 

Here, we should note the following: \textit{i)} since the antenna size increases with increasing $G_T$, the $G_T$ vs. $P_T$ balance must be maintained regarding what a UAV can physically accommodate, %(in terms of antenna dimensions),
\textit{ii)} power transfer should be administrated at a low frequency ($f$, preferably sub-GHz), as the power received by sensors ($P_R$) is inversely proportional to the square of $f$, i.e., $P_R\varpropto 1/f^2$, \textit{iii)} since the UAV has a fixed budget for power transfer ($E_{PT}$), decreasing $P_T$ means a longer $t_{PT}$, which may affect the service probability of the UAV -from (\ref{eq:1}). %i.e., an undesirable consequence. %i.e. 
The lengthened coverage lifetime, $t_{PT}$, %. Although this 
despite sounding attractive, will alter the duty cycle of sensors, %the w-p\textit{D}s will accordingly be altered, 
which cannot be tolerated always due to the certain reporting frequency requirements of IoT applications \cite{cetinkaya2017}.
Thus, these trade-offs must be carefully considered during the system design to maximize the performance metric defined by the application.

\subsection{Trip Power Consumption}
\begin{figure}[!b]
%\vspace{-2mm}
	\centering
	\begin{subfigure}{0.49\linewidth}
		\includegraphics[width=0.98\textwidth]{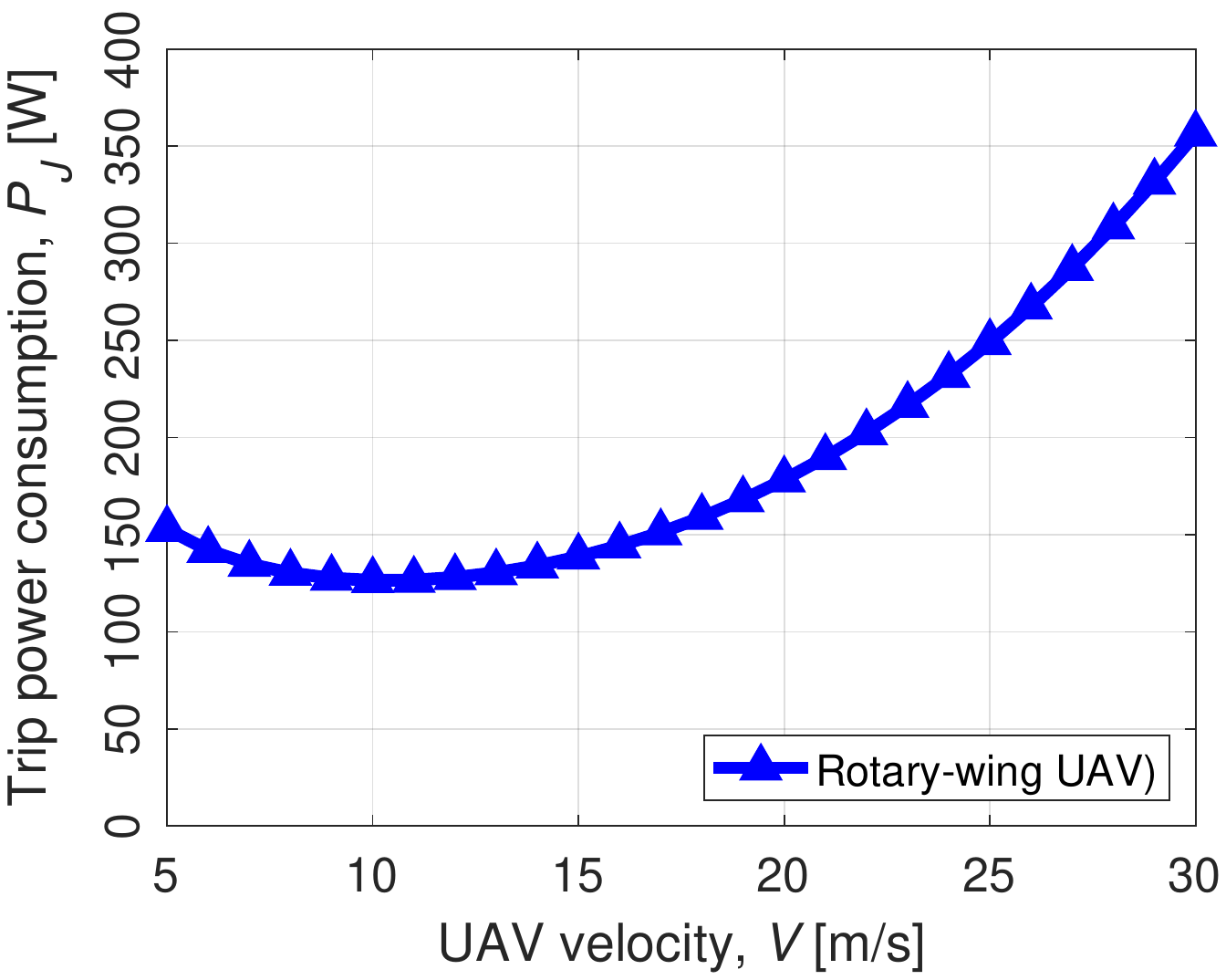}
    %\vspace{-1mm}
		\caption{}
	\end{subfigure}
	\begin{subfigure}{0.49\linewidth}
		\includegraphics[width=0.98\textwidth]{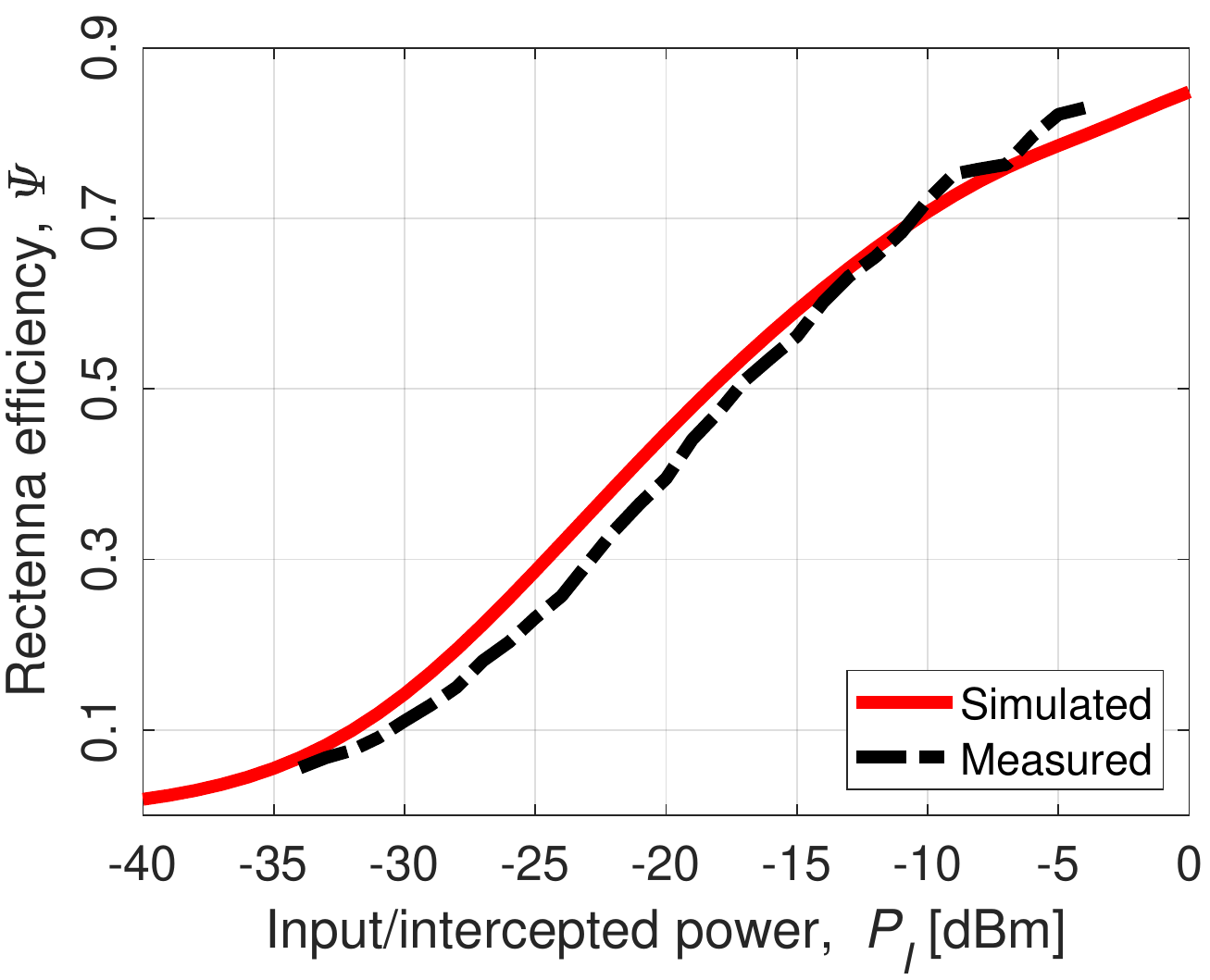}
    %\vspace{-1mm}
		\caption{}
	\end{subfigure}
	%\vspace{-3mm}
	\caption{(a) Trip power consumption, $P_J$, vs. UAV velocity, $V$,~as modelled in \cite{zeng2019energy}; (b) rectenna efficiency, $\Psi$, as a function of input/intercepted power,~$P_{I_{\chi}}$, for the rectenna in \cite{wagih2020high} ($@f_c\!=\!868$MHz).}  
	\label{fig:data_used}
\end{figure}

The rotary-wing type UAVs that we have need fixed power during their trip, which can be approximated as \cite{zeng2019energy}: 
%The fixed power that the UAV needs during travelling can be expressed as 
%\begin{equation}
%\label{eq:hovering}
%P_J = P_0 \bigg ( 1 + \frac{3 V^2}{U^2_{tip}}\bigg ) + \frac{P_i v_0}{V} + \frac{1}{2} d_0 \rho s A V^3, 
%\end{equation}
\begin{equation}
\label{eq:tripP}
%P_J(V)\!=\!P_0 \bigg ( 1 + \frac{3 V^2}{U^2_{tip}}\bigg ) + P_i \bigg (\!\sqrt{1\!+\!\frac{V^4}{4 v_0^2}}-\frac{V^2}{2 v_0^2} \bigg )^{\!\tfrac{1}{2}}\!+ \frac{d_0 \rho s A V^3}{2},
P_J(V) \approx P_0 \bigg (1 + \frac{3 V^2}{U^2_{tip}}\bigg ) + \frac{P_i v_0}{V} + \frac{1}{2} d_0 \rho s A V^3, 
\end{equation}
where $U^2_{tip}$ is the tip speed of the rotor blade, $v_0$ is the mean rotor-induced velocity when hovering, $d_0$ is the fuselage drag ratio, $\rho$ is the air density, $s$ is the rotor solidity, $A$ is the rotor disc area, %$V$ is the UAV velocity, 
and $P_0$ and $P_i$ are the UAV's blade profile power and induced power in hovering status, respectively. Here, %Note that $P_h$ refers to the case when $V\!=\!0$\ in (\ref{eq:hovering}), i.e., $P_h=P_J(V\!=\!0)=P_0+P_i$.
$P_h$ can be defined as the sum of $P_0$ and $P_i$, i.e.,:
\begin{equation}
\label{eq:P_h}
P_h = \underbrace{\frac{\delta}{8} \rho s A \Omega^3 R^3}_\text{$\triangleq P_0$} + \underbrace{(1+k) \frac{W^{3/2}}{\sqrt{2 \rho A}}}_\text{$\triangleq P_i$},
\end{equation}
where $\Omega$ is the blade angular velocity, $R$ is the rotor radius, $k$ is the incremental correction factor to induced power, and $W$ is the UAV weight. Using the respective values of each parameter given in \cite{zeng2019energy}, $P_J$ as a function of $V$ is illustrated in Fig.~{\ref{fig:data_used}}(a), which is also used in our calculations.

From (\ref{eq:2}) and (\ref{eq:tripP}), the energy that the UAV needs for~a~round trip, $E_J$, is $t_J\,\times\,P_J\,=\,\tfrac{2(R_\Delta+h_l)}{V}P_J$, where each leg consumes the half, i.e., $E_J/2$ for travelling to or from the service~point. 
%Here, the energy required for one leg of the trip, i.e., travelling to or from~the service point, is the half of $E_J$. 
%travelling to or from the charging station can be given as:
%\vspace{-2mm}
%\begin{equation}
%\label{eq:travelenergy}
%\begin{split}
%E_J\!=\!\underbrace{\frac{2d_{\Delta}}{V}}_{t_J}\overbrace{\bigg [\!P_0 \bigg (\!1\!+\!\frac{3V^2}{U^2_{tip}}\bigg )\!+\!\frac{P_i v_0}{V}\!+\!\frac{1}{2} d_0 \rho s A V^3 \bigg ]}^{P_J}.
%\end{split}
%\end{equation}
That is important, as %the UAV should spare enough energy, $E_J/2$, to make it to the nearest charging station before it depletes its battery.
the energy left in $B_{\mathrm{UAV}}$ after providing service must be enough for UAV to make it to the nearest charging station before depleting its battery, i.e., $\!E_\mathrm{left}\!\geq\!E_J/2$.
In our analyses, %we precisely pick the value of $V$ as minimizing $E_J$, referred to as $V_{\mathrm{opt}}$, so that the UAV has more resources for \textit{service}.
we evaluate the effect of $V$ in minimizing $E_J$, ensuring that the UAV has more resources for \textit{service}.

\subsection{Sensor Association}

We assume that each event area, $\Delta$, is serviced by only~one UAV at a time, i.e., each sensor is associated with one~UAV hovering at the center of $\Delta$ that the sensor falls into. Otherwise, the sensor has no service, which refers to the outage. %Note that service corresponds to $i)$ power transfer to sensors in the hotspot center; $ii)$ collection of their data. %Throughout this letter, we focus our analysis on a randomly selected sensor inside the hotspot, which is referred to as the reference sensor.

%\begin{figure}[h!]
%    \centering
%    \includegraphics[width=\linewidth]{figs/UserAssociation.JPG}
%    \caption{\textcolor{red}{Has to be rewritten.}}
%    \label{fig:2}
%\end{figure}

The power intercepted by the sensor, $P_I$, after the UAV initializes the power transfer process is: 
\begin{equation}
\label{eq:RecPow}
P_{I_{\chi}} = \frac{\mathrm{EIRP^+} G_R G_h}{\mathrm{PL}_{\chi}},
\end{equation}
where $\mathrm{EIRP^+}\!$ %\textit{effective isotropic radiated power}, 
is equal to $P_T\,\times\,G_T$, %i.e., the product of transmit power and antenna gain of the UAV, 
$G_R$ is the gain of~sensor antenna, $G_h$ is exponentially distributed fading power coefficient, $\mathrm{PL}$ is path loss as a function of distance between the UAV and sensor, %We should note that, in theory, $\mathrm{PL}$ is maximum for any sensor on the circle encapsulating the event area, i.e., when the distance is $R_{RF}$ -worst case scenario. For any other sensor in the event area, the UAV-to-sensor distance is $d_{3D}$, which refers to a lower $\mathrm{PL}$. The minimum PL is achieved at the center of the event area, i.e., when the distance is $h_{UT}$, which refers to the best case scenario. 
and finally, $\chi$ refers to $\mathrm{LoS}$ and $\mathrm{NLoS}$, i.e., the line of sight (LoS) and Non-LoS, indicating the condition of the air-to-ground link between the UAV and sensors. %pair. 
%The sensors use $P_{R_{\chi}}$ to make some readings and deliver their findings to the UAV powering them. 

To perform sensory operations, however, $P_{I_{\chi}}$ has to be converted into utilizable DC power, $P_{R_{\chi}}$, using a rectifying antenna or \textit{rectenna}. %Usually, $P_{out}$ is a non-linear function of $P_{in}$, due to the non-linearity of components used in the EH circuitry, such as diode(s). 
%The RF-to-DC conversion efficiency $\eta$ ($0\!\leq\!\eta\!<\!1$) often increases with increasing $P_{in}$; however, this increase is typically non-linear due to the components used in the EH circuitry, such as diodes. In practice, the rectenna starts operating after $P_{in}$ reaches up the minimum power threshold $P_{th}$, i.e., the \textit{sensitivity}, and gives  a constant $P_{out}$ when $P_R$ exceeds the saturation power threshold $P_{sat}$ of the EH circuitry \cite{RF_survey}. Based on these behaviors, the function $P_{out}(P_{in})\!=\!\eta(P_{in})\!\cdot\!P_{in}$ should follow an ``S"-like shape (Sigmoid curve), and to model this shape, use of logistic function is proposed by the previous works \cite{boshkovska2015, wang2017}. %The rectenna efficiency can be alternatively defined as a logistic function \cite{bibid}; 
%However, \cite{alevizos2018a} revealed that the models based on this assumption are erroneous, and accordingly came up with the following piecewise linear function, where $P_{out}(P_{in})$ is captured as a high-order polynomial as:
Considering the sensitivity and saturation of rectennas, the research field has agreed on the following piecewise linear function, capturing $P_{R_{\chi}}$ as a high-order polynomial~\cite{alevizos2018nonlinear}:
\begin{equation}
P_{R_{\chi}}(P_{I_{\chi}}) \triangleq	
\begin{cases}
0,&~P_{I_{\chi}} \in [0,P_{th}),\\
%\eta(P_R)\!\cdot\!P_R,&~P_R \in [P_{th}, P_{sat}],\\
(p_w + \sum\limits_{j=1}^{w-1} p_j P_{I_{\chi}}^{w-j})\!\cdot\!P_{I_{\chi}},&~P_{I_{\chi}} \in [P_{th}, P_{sat}],\\
P_{R_{\chi}}(P_{sat}),&~P_{I_{\chi}} \geq P_{sat},
%\eta\!(P_{sat})\,P_{sat},&~x \geq P_{sat},
\end{cases}
\label{eq:rectenna_efficiency}
\end{equation}
where $P_{R_{\chi}}(P_{I_{\chi}})$ is non-decreasing and continuous for all $P_{I_{\chi}}\!\in\!\mathbb{R},\,P_{I_{\chi}}\!\geq\!0$. In our analyses, the rectennas are assumed to operate in the ideal region, i.e., $P_{I_{\chi}}\!\in\![P_{th}, P_{sat}]$ for all sensors. Hence, $(p_w\!+\!\sum_{j=1}^{w-1} p_j P_{I_{\chi}}^{w-j})$ is the rectenna (or RF-to-DC conversion) efficiency as a function of $P_{I_{\chi}}$, i.e.,
$\Psi(P_{I_{\chi}})$ ($0\!\leq\!\Psi(P_{I_{\chi}})\!<\!1$), with $w$ being the degree of polynomial and $\{p_j\}_{j=1}^{w}$ the respective coefficients. Furthermore, $\Psi(P_{I_{\chi}})$ is calculated using real data outsourced from \cite{wagih2020high}, where Fig.~\ref{fig:data_used}(b) depicts the measured and simulated behaviors of the rectenna design of authors.

%Here, $P_{in}$ and $P_{out}(P_{in})$ are in Watt, and $P_{out}(P_{in})$ is non-decreasing and continuous for all $P_{in}\!\in\!\mathbb{R},\,P_{in}\!\geq\!0$. %The coefficients $\{p_j\}_{j=1}^{w}$ can be obtained by using standard curve fitting tools for any empirical dataset, which will give a mathematically tractable expression of $P_{out}(P_{in})$ with sufficient precision.
%In this work, the rectenna is assumed to operate in the ideal region, i.e., $P_{in}\!\in\![P_{th}, P_{sat}]$ for all w-p\textit{T}s.

$\mathrm{PL_{LoS}}$ and $\mathrm{PL_{NLoS}}$ in (\ref{eq:RecPow}) can be given as \cite{li2019joint}:
%\begin{equation}
\begin{align}
    \mathrm{PL_{LoS}}=&20\mathrm{log_{10}}\bigg (\frac{4\pi f_c d_\mathrm{{3D}}}{c}\bigg )+\eta_{\mathrm{LoS}},\notag\\
    \mathrm{PL_{NLoS}}=&20\mathrm{log_{10}}\bigg (\frac{4\pi f_c d_{\mathrm{3D}}}{c}\bigg )+\eta_{\mathrm{NLoS}},
    %\mathrm{PL_{LoS}}=&\mathrm{max}(23.9-1.8\mathrm{log}_{10}(h_{u}),20)\\&\times\mathrm{log}_{10}{(d_{3D})}+20\mathrm{log}_{10}\big (\tfrac{40\pi f_c}{3}\big ),\notag\\
%\end{aligned} 
%\end{equation}
%begin{equation}
%\begin{aligned}
    %\mathrm{PL_{NLoS}}=&\mathrm{max}(\mathrm{PL_{LoS}},-12+(35-5.3\mathrm{log}_{10}(h_u))\\&\times\mathrm{log}_{10}(d_{3D})+20\mathrm{log}_{10}\big (\tfrac{40\pi f_c}{3}\big ),\notag
\end{align}
%\end{equation}
%both of which hold for rural macrocell scenario and for  $1.5$m$\leq h_u\leq300$m and $d_{2D}\leq10$km.
where $f_c$ is the carrier frequency, $c$ is the speed of light, and $\eta_{\mathrm{LoS}}$, $\eta_{\mathrm{NLoS}}$ are average additional loss, depending on the environment for LoS, and NLoS links, respectively. Furthermore, the probability that the UAV has a LoS air-to-ground link with a sensor %, $P_{LoS}(R_u)$, 
%for $10$m$<\!h_{UT}\!\leq\!40$m is 
can be formulated as \cite{liao2021hotspot}:
\begin{equation}
P_{\mathrm{LoS}}=\frac{1}{1+\gamma \mathrm{exp}(-\delta(90-\theta_B/2-\gamma))},    
\end{equation}
%\begin{align*}
%    P_{LoS} &=
%    \begin{cases}
%       1, &d_{2D}\leq d_1\\
%       \frac{d_1}{d_{2D}} + \mathrm{exp} \big ( \frac{-d_{2D}}{p_1}\big ) \big ( 1-\frac{d_1}{d_{2D}}\big ), &d_{2D}>d_1
%    \end{cases}\numberthis
%\end{align*}
%where $p_1=\mathrm{max}(15021\mathrm{log}_{10}(h_u)-16053,1000)$ and $d_1=\mathrm{max}(1350.8 \mathrm{log}_{10}(h_u)-1602,18)$. For $h_u$ exceeding $40$m (up to $300$m), $P_{LoS}$ becomes $100\%$. 
where $\gamma$ and $\delta$ are constant values that depend on the environment (e.g., suburban, high-rise urban), and ($90-\theta_B/2$) is the elevation angle of the UAV. Finally, the probability of NLoS is always $P_{\mathrm{LoS}}=1-P_{\mathrm{NLoS}}$.

%\color{blue}
%Shadow fading also depends on the LoS/NLoS condition. Standard deviation of the shadow fading for LoS condition is $\sigma_{\mathrm{LoS}} = 4.64\mathrm{exp}(−0.00066h)$, and for NLoS condition, $\sigma_{\mathrm{NLoS}} = 6$ \cite{3GPP}. Finally, the channel is assumed to exhibit Rayleigh block fading characteristics.
%\color{black}

%According to [12], the probability that the UAV has a LoS channel to the reference user is given as:
%\begin{equation}
%P_l(R_u) = \frac{1}{1+a \exp(-b(\frac{180}{\pi}\arctan(\frac{h}{\sqrt{R^2_u-h^2}})-a))},
%\end{equation}
%where $a$ and $b$ are constants that related to the environment, and $h$ is the altitude of the UAV. Moreover, the probability of NLoS is $P_n(R_u)=1-P_1(R_u)$.

%\begin{definition}[Coverage probability]
%\label{def:CovProb}
Based on the equations given above, we can define the coverage probability of the UAV conditioned on $R_\Delta$ as:
\begin{equation}
\label{eq:CovProb1}
    P_{\mathrm{cov|R_\Delta}} = P_{\mathrm{(e|R_\Delta)}}P_{\mathrm{cov,s}},
\end{equation}
where $P_{\mathrm{(e|d_{\Delta})}}$ is given in (\ref{eq:1}). Hence, the unconditional coverage probability can be expressed as:
\begin{equation}
\label{eq:CovProbAc}
    P_\mathrm{cov} = P_\mathrm{e} P_{\mathrm{cov,s}},
\end{equation}
in which:
\begin{equation}
\label{eq:CovProb}
    P_{\mathrm{cov,s}} = \mathbb{P}%\Big ( \frac{P_{R_{\chi}}}{\sigma^2}\geq \beta \Big),
    (P_{R_{\chi}}\geq \Gamma_{th}),
\end{equation}
where %$\sigma^2$ is the noise power, and $\beta$ is the signal-to-noise-ratio (SNR) ratio
$\Gamma_{th}$ is the minimum power that has to be received by a sensor to become active, referring to the sensitivity, probe its vicinity, and deliver the data it collects to the UAV. Here, we should note that $d_{3D}$, and so $R_{RF}$, can be assumed as equal to $h_{UT}$ due to directivity ($h_{UT}\!=\!\cos{(\theta/2)}\!\times\!R_{RF}$ actually, but $\theta/2$ is quite small; hence, $R_{RF}\!\approx\!h_{UT}$). That leads to the assumption that $P_{R_\chi}$ will be equal for each point in the event area, i.e., all sensors in $\Delta$ will receive the same power irrespective of their locations. Therefore, $P_{R_\chi}$ does not need to be averaged when $P_{\mathrm{cov,s}}$ is calculated.

\begin{comment} 
When the user associates with the UAV, the received power is:
\begin{align*}
    p_u &=
    \begin{cases}
       p_l = \eta_l \rho_u G_l R_u^{-\alpha_l}, &\textnormal{in case of LoS,}\\
       p_n = \eta_n \rho_u G_n R_u^{-\alpha_n}, &\textnormal{in case of NLoS,}
    \end{cases}
\end{align*}
where $\rho_u$ is the transmission power of the UAV, $R_u$ is the distance between the reference user and the UAV, $\alpha_l$ and~$\alpha_n$ are the path-loss exponents, $G_l$ and $G_n$ are the fading gains that follow gamma distribution with shape and scale parameters $(m_l, \frac{1}{m_l})$ and $(m_n,\frac{1}{m_n})$, $\eta_l$ and $\eta_n$ denote the mean additional losses for LoS and NLoS transmission, respectively.
\end{comment}

%\end{definition}

%\begin{figure}[h!]
%    \centering
%    \includegraphics[width=\linewidth]{figs/CoverageProb.JPG}
%    \caption{\textcolor{red}{Has to be rewritten.}}
%    \label{fig:2}
%\end{figure}

\section{Performance Metrics}
\label{sec:PA}
In this section, we first derive the service probability conditioned on the distance to the nearest charging station, i.e., $P_{(\mathrm{e|R_\Delta})}$, to calculate the unconditioned service probability,~$P_\mathrm{e}$. Then, we find $P_{\mathrm{cov,s}}$, and hence, study the coverage probability, $P_{\mathrm{cov}}$, for the envisioned energy-neutral network scenario.

\subsection{Service Probability}
%\textcolor{red}{In this subsection, we analyze the statistics of the availability probability of the UAV. This analysis will be used to study the coverage probability in the next subsection.}

By substituting for (\ref{eq:2}) in (\ref{eq:1}), we can derive the service probability given the value of $R_\Delta$ as:
%\begin{lemma}[Conditional %Availability Probability] Given the value of $R_\Delta$, the availability probability is given by:
\begin{equation}
\label{eq:ServProb}
    P_{(\mathrm{e|R_\Delta})} = \frac{\zeta-2 (R_\Delta+h_l)P_J}{\zeta-2 (R_\Delta+h_l)(P_J-P_h)+Vt_{Ch}P_h},
    %P_{(\mathrm{e|R_\Delta})} = \frac{\zeta-2 P_J(R_\Delta+h_l)}{\zeta+VP_h(t_{PT}+t_{CH})+2(R_\Delta+h_l)(P_h-P_J)},
    %P_{(\mathrm{e|d_{\Delta}})} = \frac{\zeta-2 d_{\Delta}P_J}{\zeta-2 d_{\Delta}(P_J-P_h)+Vt_{Ch}P_h}
\end{equation}

\noindent where \small$\zeta = VB_{\mathrm{UAV}}-Vt_{PT}P_T$\normalsize.
%\end{lemma}
%\begin{proof}
%The above result follows directly by substituting for (\ref{eq:2}) in (\ref{eq:1}).
%\end{proof}
It should be noted that (\ref{eq:ServProb}) only holds if %\small$d_{\Delta}\leq$\normalsize $~\tfrac{VB_{\mathrm{UAV}}}{2P_J}$; otherwise, \small$P(\mathrm{e|d_{\Delta}})\!=\!0$\normalsize.
\small$R_\Delta\leq$\normalsize $~\tfrac{VB_{\mathrm{UAV}}-2h_l P_J}{2P_J}$; otherwise, \small$P(\mathrm{e|R_\Delta})\!=\!0$\normalsize. If this condition is
not satisfied, it means that $B_{\mathrm{UAV}}$ is not large enough to support the energy required for the round trip. Hence, there will not be enough power for the UAV to provide
service in $\Delta$. In addition, when \small$R_\Delta\!=\!0$\normalsize, $h_l$ is also $0$, because the UAV cannot descent when it is still on the recharging station. %, i.e.,, which is the case if $R_{\Delta}=0$.
In that case, the maximum service probability is achieved, i.e., \small$P_{(\mathrm{e|R_\Delta\,=\,0})}$\normalsize $~=\tfrac{\zeta}{\zeta+Vt_{CH}P_h}$. 
%Here, we should note that, when $R_{\Delta}=0$, $h_l$ is also $0$, because the UAV cannot descent when it is still on the recharging station, which is the case when $R_{\Delta}=0$, and this explains the value of \small$P_{(\mathrm{e|R_\Delta\!=\!0})}$.

Now, using (\ref{eq:ServProb}), let's calculate the CDF of the conditional service probability, \small$F_{P_{(\mathrm{e}|\mathrm{R_\Delta})}}(x)$\normalsize, as:
\begin{align*}
&F_{P_{(\mathrm{e}|\mathrm{R_\Delta})}}(x)\\
&= \mathbb{P}(P_{(\mathrm{e}|\mathrm{R_\Delta})}\leq x)\\
&= \mathbb{P}\bigg ( \frac{\zeta-2 (R_\Delta+h_l)P_J}{\zeta-2 (R_\Delta+h_l)(P_J-P_h)+Vt_{Ch}P_h}\leq x \bigg),
\end{align*}

\noindent and given that $P_{(\mathrm{e}|\mathrm{R_\Delta})}$ is a decreasing function of $R_\Delta$, the preimage can be obtained as:
\begin{equation}
\begin{split}
%F_{P_{(\mathrm{e}|\mathrm{d_{\Delta}})}}(x)
%= \mathbb{P}\bigg (d_{\Delta} \geq \frac{\zeta (1-x)-xVt_{Ch}P_h}{2(1-x)P_J+2xP_h} \bigg ).    
= \mathbb{P}\bigg (R_\Delta \geq \frac{\zeta (1-x)-h_l\kappa-xVt_{Ch}P_h}{\kappa} \bigg ),
\end{split}
\end{equation}

\noindent where $\kappa = 2[P_J(1-x)+xP_h]$.

Hence, the CDF becomes:
%\begin{lemma}[CDF of Conditional Availability Probability]
%The CDF of the conditional availability probability is given by:
\begin{equation}
\label{eq:CDFofPaRs}
F_{P_{(\mathrm{e|R_\Delta})}}(x)=e^{-\lambda_{Ch}\pi Q^2(x)}, \end{equation}

\noindent in which:
\begin{equation*}
%Q(x)=\frac{\zeta V(1-x)-xVP_s(t_{PT}+t_{CH})-\kappa h_l}{\kappa}.
Q(x)=%\frac{\zeta (1-x)-xVt_{Ch}P_h}{2(1-x)P_J+2xP_h}.
\frac{\zeta (1-x)-h_l\kappa-xVt_{Ch}P_h}{\kappa}.
\end{equation*}
%\begin{equation}
%0\le x \le \frac{\zeta -2h_lP_m}{\zeta +VP_s(t_{PT}+t_{CH})+2h_l(P_s-P_m)},
%\end{equation}
%\end{lemma}
%\begin{comment}

\noindent Since the minimum value of $R_\Delta = 0$, and its maximum value for a non-zero availability probability is $\tfrac{VB_{\mathrm{UAV}}-2h_lP_J}{2P_J}$, then:
\small
\begin{equation*}
%0\le x \le \frac{B_{\mathrm{UAV}}}{B_{\mathrm{UAV}}+t_{Ch}P_h}.
0\le x \le \frac{\zeta}{\zeta+Vt_{CH}P_h}
\end{equation*}
\normalsize
%\end{comment}

\begin{comment}
\begin{proof} Proof of Lemma 2 is as follows:
\begin{align*}
&F_{P_{(\mathrm{a}|\mathrm{R_\Delta})}}(x)\\
&= \mathbb{P}(P_{(\mathrm{a}|\mathrm{R_\Delta})}\leq x)\\
&= \mathbb{P}\bigg ( \frac{\zeta-2 P_m(R_\Delta+h_l)}{\zeta+VP_s(t_{PT}+t_{CH})+2(R_\Delta+h_l)(P_s-P_m)}\leq x \bigg) \numberthis.
\end{align*}
Given that $P_{(\mathrm{a}|\mathrm{R_\Delta})}$ is a decreasing function of $R_\Delta$, the preimage can be obtained as:\\
$F_{P_{(\mathrm{a}|\mathrm{R_\Delta})}}(x)
= \mathbb{P}\bigg (R_\Delta \geq \frac{\zeta V(1-x)-xVP_s(t_{PT}+t_{CH})-\kappa h_l}{\kappa} \bigg )$.\\
Given that the minimum value of $R_\Delta = 0$, and its maximum value for a non-zero availability probability is $\tfrac{\zeta-2P_mh_l}{2P_m}$, then:\vspace{2mm}\\
\[0\le x \le \frac{\zeta -2h_lP_m}{\zeta +VP_s(t_{PT}+t_{CH})+2h_l(P_s-P_m)}.\]
\end{proof}
\end{comment}

%In the following theorem, we derive the availability probability. 

Using these results, we can find the service probability as:
\begin{align*}
\label{eq:P_e}
    %\[
    P_\mathrm{e} &= \mathbb{E}_{\Phi_{Ch}} [P_{(\mathrm{e}|\mathrm{d_{\Delta}})}] \\&= \int_{0}^{\infty} (1-F_{P_{(\mathrm{e}|\mathrm{d_{\Delta}})}}(x)) \,\mathrm{d}x\\&=
    \int_{0}^{\tfrac{\zeta}{\zeta + Vt_{Ch}P_h}} (1-e^{-\lambda_{Ch}\pi Q^2(x)}) \,\mathrm{d}x.\numberthis
    %\]
\end{align*}
%where $x_{\mathrm{min}}$ and $x_{\mathrm{max}}$ can be found by using the minimum and maximum values of $R_\Delta$. 

\begin{comment}
\begin{theorem}Availability Probability. The availability probability of the UAV $P_a$ is:
\[P_a =  \int_{0}^{\frac{\zeta-2P_mh_l}{2P_m}} e^{-\lambda_{Ch}\pi Q(x)^2} \,\mathrm{d}x. \]\vspace{-3mm}
\end{theorem}
\begin{proof}
The above expression follows by substituting the results in Lemma 2 into:
\[P_\mathrm{a} = \mathbb{E}_{\Phi_c} [P_{(\mathrm{a}|\mathrm{R_\Delta})}] = \int_{0}^{\infty} 1-F_{P_{(\mathrm{a}|\mathrm{R_\Delta})}}(x) \,\mathrm{d}x. \]\vspace{-3mm} 
\end{proof}
\end{comment}

\subsection{Coverage Probability}
Based on the equations derived in the previous subsections,
we can now work on the coverage probability given in
(\ref{eq:CovProb}). First, we need to reexpress $P_{\mathrm{cov,s}}$ as follows:
%\vspace{-2mm}
%\begin{lemma}[Coverage Probability] The coverage probability of the UAV from (\ref{eq:CovProb}) can be given as:%\vspace{2mm}\\
\small
\begin{align*}
\label{eq:P_cov_s}
%\scriptsize
&\hspace{-3mm}P_{\mathrm{cov,s}} \\
&\hspace{-3mm}\!=\! P_{{\mathrm{cov}}_{\mathrm{LoS}}} P_\mathrm{LoS} + P_{\mathrm{{cov}_{NLoS}}} P_\mathrm{NLoS}\\%\vspace{1mm}\\
&\hspace{-3mm}\!=\!\mathbb{P}\big(P_{R_\mathrm{LoS}}\!\geq\!\Gamma_{th}\big)P_\mathrm{LoS}+\mathbb{P}\big(P_{R_\mathrm{NLoS}}\geq\Gamma_{th}\big)P_\mathrm{NLoS}\\%\vspace{2mm}\\
&\hspace{-3mm}\!=\!\mathbb{P}\big(\eta_{\mathrm{LoS}}\tfrac{\mathrm{EIRP\!^+}G_RG_h}{\mathrm{PL_{LoS}}}\geq\Gamma_{th}\big)P_\mathrm{LoS}+...\\&\hspace{29mm}...+\mathbb{P}\big(\eta_{\mathrm{NLoS}}\tfrac{\mathrm{EIRP\!^+}G_RG_h}{\mathrm{PL_{NLoS}}}\geq\Gamma_{th}\big)P_\mathrm{NLoS}\\
&\hspace{-3mm}\!=\!\mathbb{P}\bigg(\!G_h\!\geq\!\tfrac{\Gamma_{th} \mathrm{{PL_{LoS}}}}{^{\!\eta_{\mathrm{LoS}}\mathrm{EIRP\!^+}\!G_R}}\!\bigg)\!P_\mathrm{\mathrm{LoS}}\!+\!\mathbb{P}\bigg(\!G_h\!\geq\!\tfrac{\Gamma_{th} \mathrm{{PL_{NLoS}}}}{^{\eta_{\mathrm{NLoS}}\mathrm{{EIRP\!^+}}\!G_R\!}}\!\bigg)\!P_\mathrm{\mathrm{NLoS}}\\%\vspace{2mm}\\
&\hspace{-3mm}\!=\!e^{-\Big(\frac{\Gamma_{th} \mathrm{{PL_{LoS}}}}{\eta_{\mathrm{LoS}}\mathrm{EIRP\!^+} G_R }\Big)}\!P_\mathrm{\mathrm{LoS}}\!+\!e^{-\Big(\frac{\Gamma_{th} \mathrm{{PL_{NLoS}}}}{\eta_{\mathrm{NLoS}}\mathrm{EIRP\!^+} G_R}\Big)}\!P_\mathrm{\mathrm{NLoS}}\numberthis.
\end{align*}
%\noindent where \textcolor{red}{$G_l$} is Rayleigh fading channel gain, which is exponentially distributed.
%\end{lemma}
\normalsize
Therefore, the coverage probability, $P_{\mathrm{cov}}$ in (\ref{eq:CovProbAc}), can be finally calculated by using (\ref{eq:P_e}) and (\ref{eq:P_cov_s}).

%%% UNUSED %%%
%\textcolor{cyan}{Due to the direct influence of service probability, $P_\mathrm{e}$, on coverage probability, $P_{\mathrm{cov}}$, and the fact that $d_{\Delta}$ varies from one $\Delta$ to another, it is important to understand how $d_{\Delta}$ impacts $P_{\mathrm{cov}}$. Hence, by substituting for $P_{\mathrm{cov,d_{\Delta}}}$ in (\ref{eq:CDFofPaRs}), we derive the complementary cumulative distribution of the coverage probability given $d_{\Delta}$ as:%CDF of the conditional coverage probability given in (\ref{eq:CovProb}).
%\begin{lemma}[Conditional CDF of Coverage Probability] The complementary cumulative distribution of coverage probability given $R_\Delta$ is:
%\begin{equation}
%F_{P_{\mathrm{cov}|\mathrm{d_{\Delta}}}}(\Theta)=1-F_{P_{(\mathrm{e}|\mathrm{d_{\Delta}})}}\big (\tfrac{\Theta}{P_{\mathrm{cov,s}}}\big ),   
%\end{equation}
%in which:
%\begin{equation*}
%0\le \Theta \le \tfrac{\zeta}{\zeta+Vt_{Ch}P_h}P_{\mathrm{cov,s}}.
%\end{equation*}
%Here, we should note that the range of values of $\Theta$ in the above result reflects the maximum and minimum achievable values of $P_{\mathrm{cov|d_{\Delta}}}$. In particular, the minimum achievable value $P_{\mathrm{cov|d_{\Delta}}} = 0$ reflects the scenario where $d_{\Delta}$ is too large that the UAV is always unavailable. On the other hand, the maximum achievable value is $P_{\mathrm{cov|d_{\Delta}}} = \tfrac{\zeta}{\zeta+Vt_{Ch}P_h}P_{\mathrm{cov,s}}$, which reflects the scenario where $d_{\Delta} = 0$.}
%\end{remark}

\vspace{4mm}
\section{Numerical Results}
\label{sec:results}

In this section, we calculate the coverage probability using the derived equations %and  \textcolor{red}{Monte-Carlo simulations} 
to study the effect of recharging time and station density, maximum battery size, and UAV speed. Unless otherwise stated, Table~\ref{tab:parameters} provides the parameter values, where $\eta_{\textrm{LoS}}$ and $\eta_{\textrm{NLoS}}$ are for high-rise urban scenario. 

\begin{table}[b!]
%\vspace{-1mm}
\caption{Parameter values.}
%\vspace{-1mm}
\label{tab:parameters}
\renewcommand*{\arraystretch}{1.3}
    \centering
    \small
    \begin{tabular}{cccc}
    \hline
        Service-related\\\hline 
        $P_T$ & $21$ [dBm] & $G_T$ & $15$ [dBi]\\ $\theta_B$ & $30.8$ [$^{\circ}$] & $f_c$ & $868$ [MHz]\\
        $c$ & $3\times10^8$ [m/s] & $\Gamma_{th}$ & $1$ [$\mu$W]\\\hline
        Environment-related &\cite{almarhabi2021lora} & & \cite{almarhabi2021lora}\\\hline
        $\eta_{\mathrm{LoS}}$ & $1.6034$ [dB] &
        $\gamma$ & $27.1157$ \\ $\eta_{\mathrm{NLoS}}$ & $29.6462$ [dB] &  $\delta$ & $0.1232$ \\\hline
        %Network parameters \textcolor{red}{\cite{bideb}}\\\hline $\sigma$ & a & $\beta$ & \\\hline
        UAV-related & & & \cite{zeng2019energy}\\\hline
        $B_{\mathrm{max}}$ & $770$ [Wh] & $P_h$ & $168.48$ [W]\\
        $V$ & $10.36$ [m/s] & $P_J$ & $126.395$ [W]\\\hline
        Others\\\hline
        $\lambda_{Ch}$ & $10^{-6}$ [m$^{-2}$] & $G_R$ & $9$ [dBi]\\
        $h_{Ch}$ & $100$ [m] & $h_l$ & $80$ [m]
        
        %\textcolor{cyan}{
        %$P_0$} & a & \textcolor{cyan}{$U_{tip}$} & b\\
        %\textcolor{cyan}{$P_i$} & a & \textcolor{cyan}{$v_0$} & b\\
        %\textcolor{cyan}{$d_0$} & a & \textcolor{cyan}{$\rho$} & b\\
        %\textcolor{cyan}{$s$} & a & \textcolor{cyan}{$A$} & b\\\hline
    \end{tabular}
    %\label{tab:my_label}
\end{table}

\begin{figure*}[!t]
%\vspace{-2mm}
	\begin{subfigure}{0.33\textwidth}
		\includegraphics[width=0.99\textwidth]{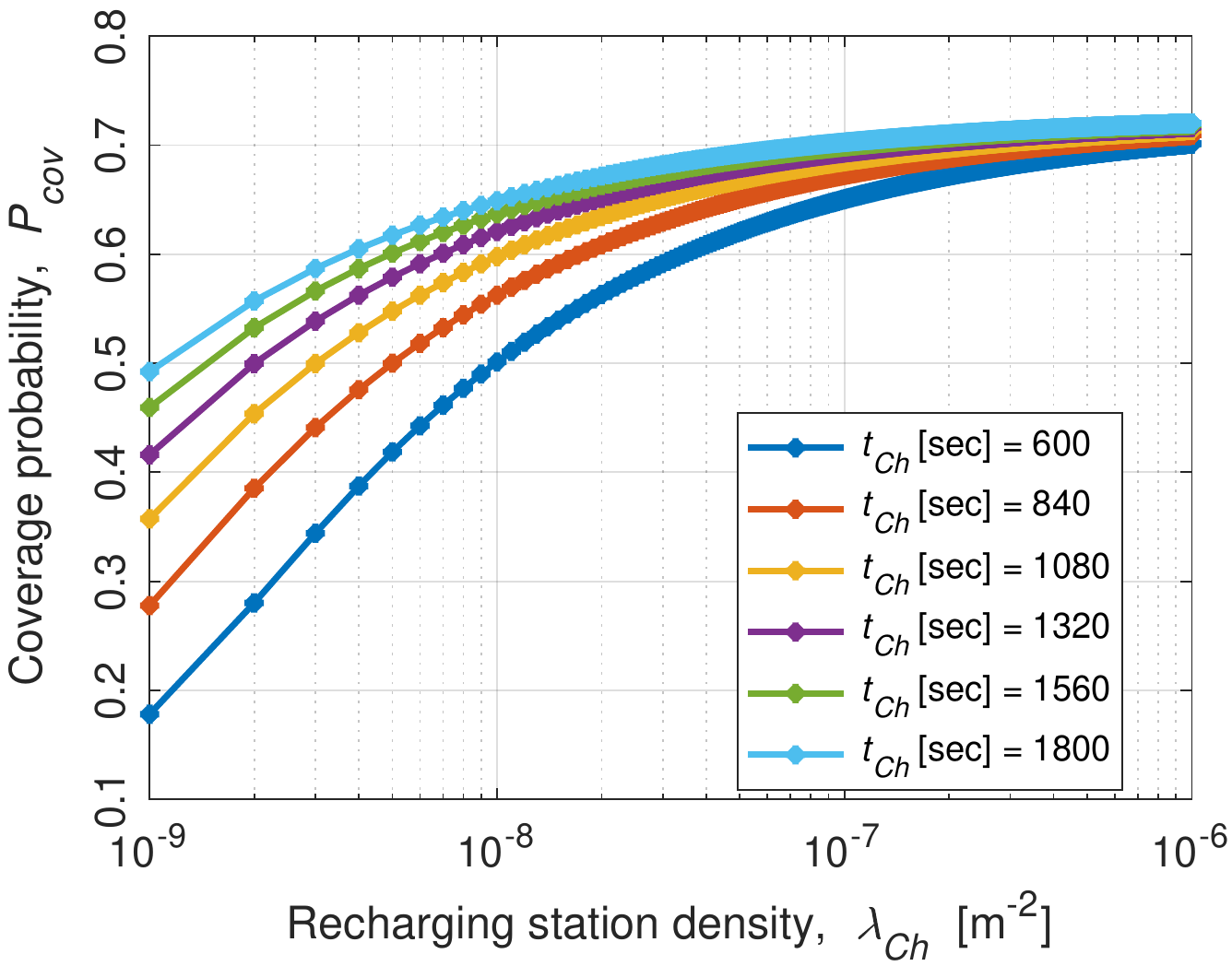}
        %\vspace{-1mm}
		\caption{}
		%\vspace{-1mm}
	\end{subfigure}
	\hspace{-2mm}
	\begin{subfigure}{0.33\textwidth}
		\includegraphics[width=0.99\textwidth]{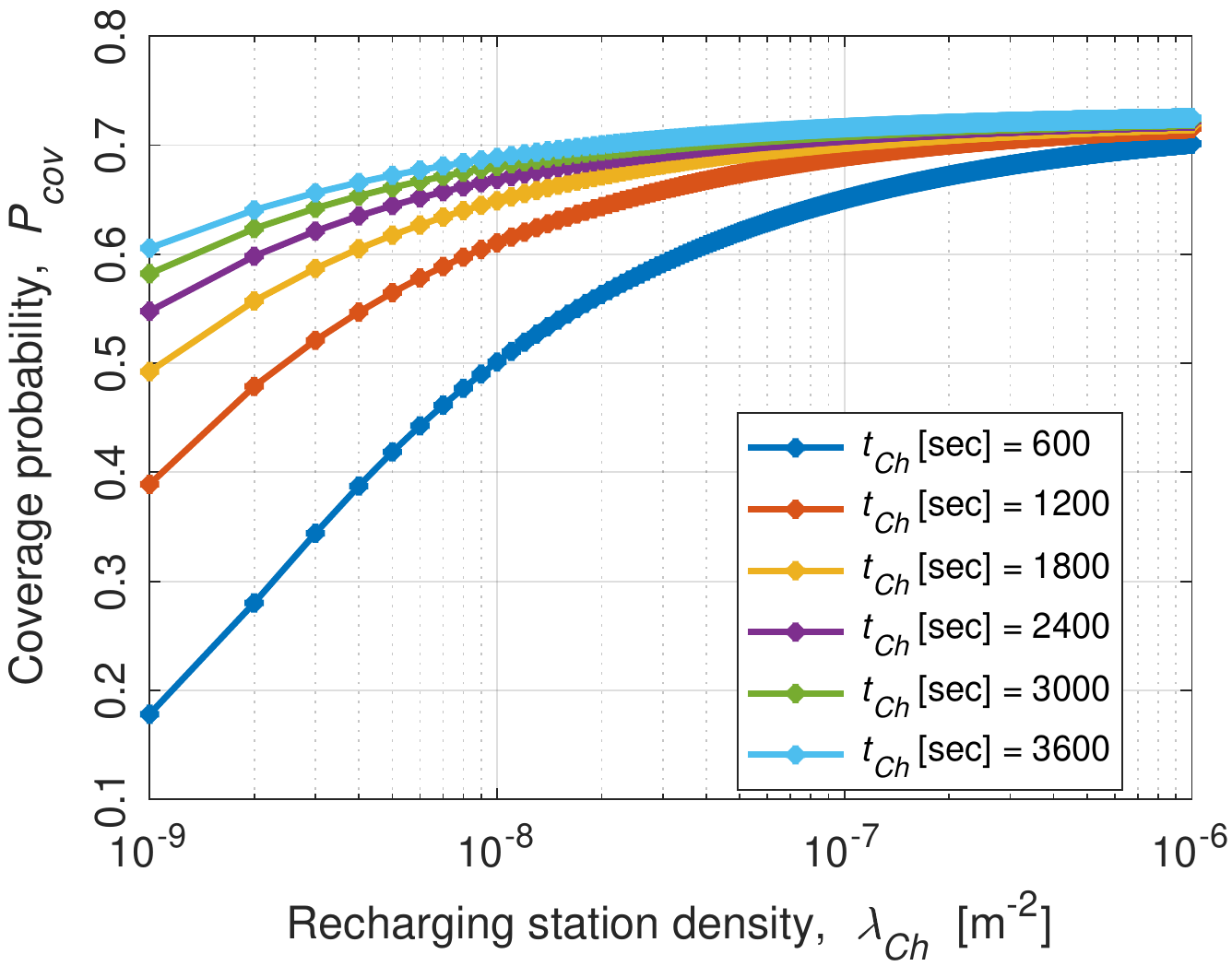}
    %   \vspace{-1mm}
		\caption{}
	%	\vspace{-1mm}
	\end{subfigure}
	\hspace{-2mm}
	\begin{subfigure}{0.33\textwidth}
		\includegraphics[width=0.99\textwidth]{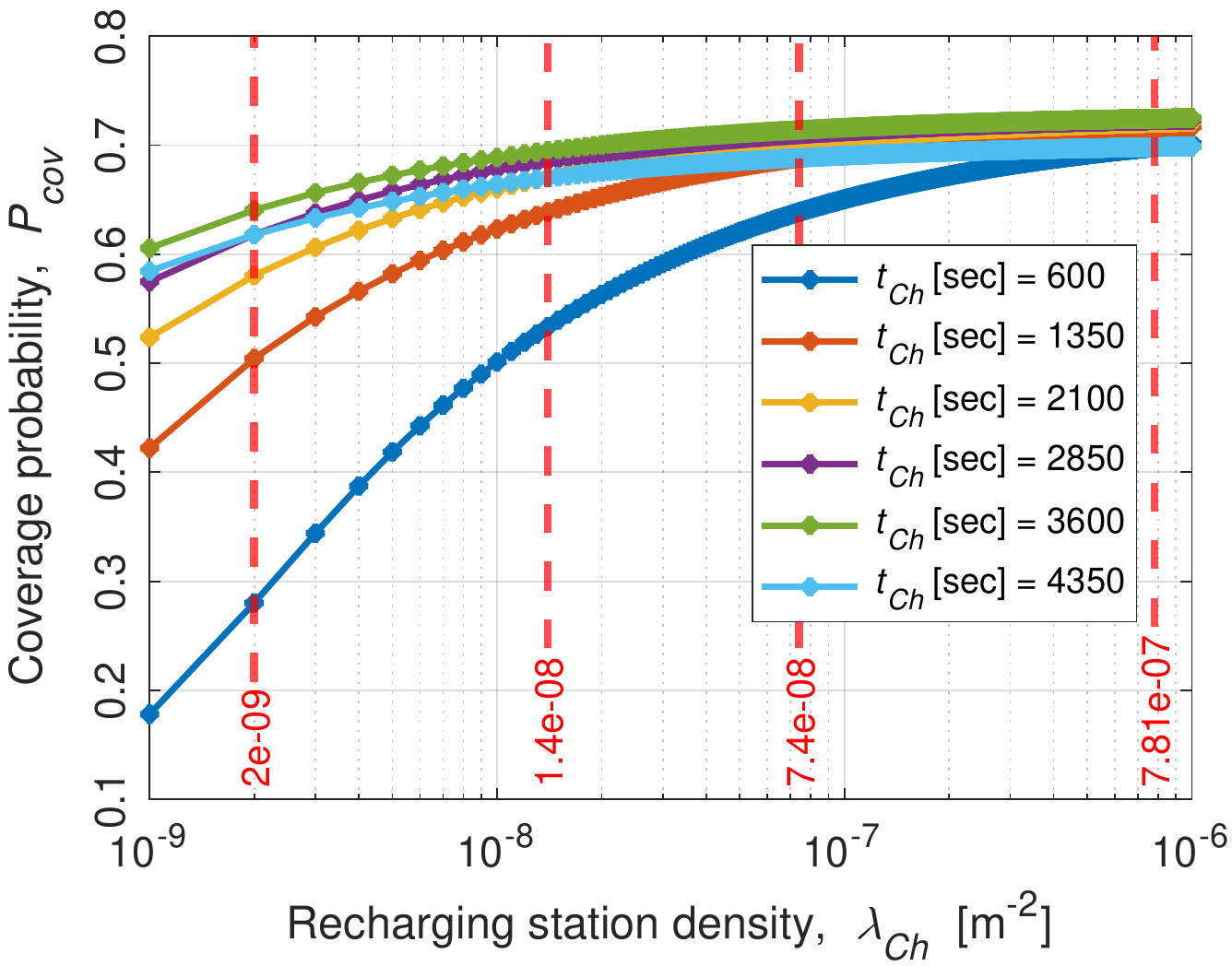}
    %\vspace{-1mm}
		\caption{}
	%\vspace{-1mm}
	\end{subfigure}
	\caption{Coverage probability, $P_{cov}$, vs. recharging station density, $\lambda_{Ch}$, for different $t_{Ch}$: (a) half-charged; (b) fully-charged; (c) overflowed.}  
	\vspace{2mm}
	\label{fig:Pcov_vs_lambda_for_tch}
\end{figure*} 

\begin{figure}[!t]
	\vspace{1mm}
	\centering
	\begin{subfigure}{0.51\linewidth}
		\includegraphics[width=0.98\textwidth]{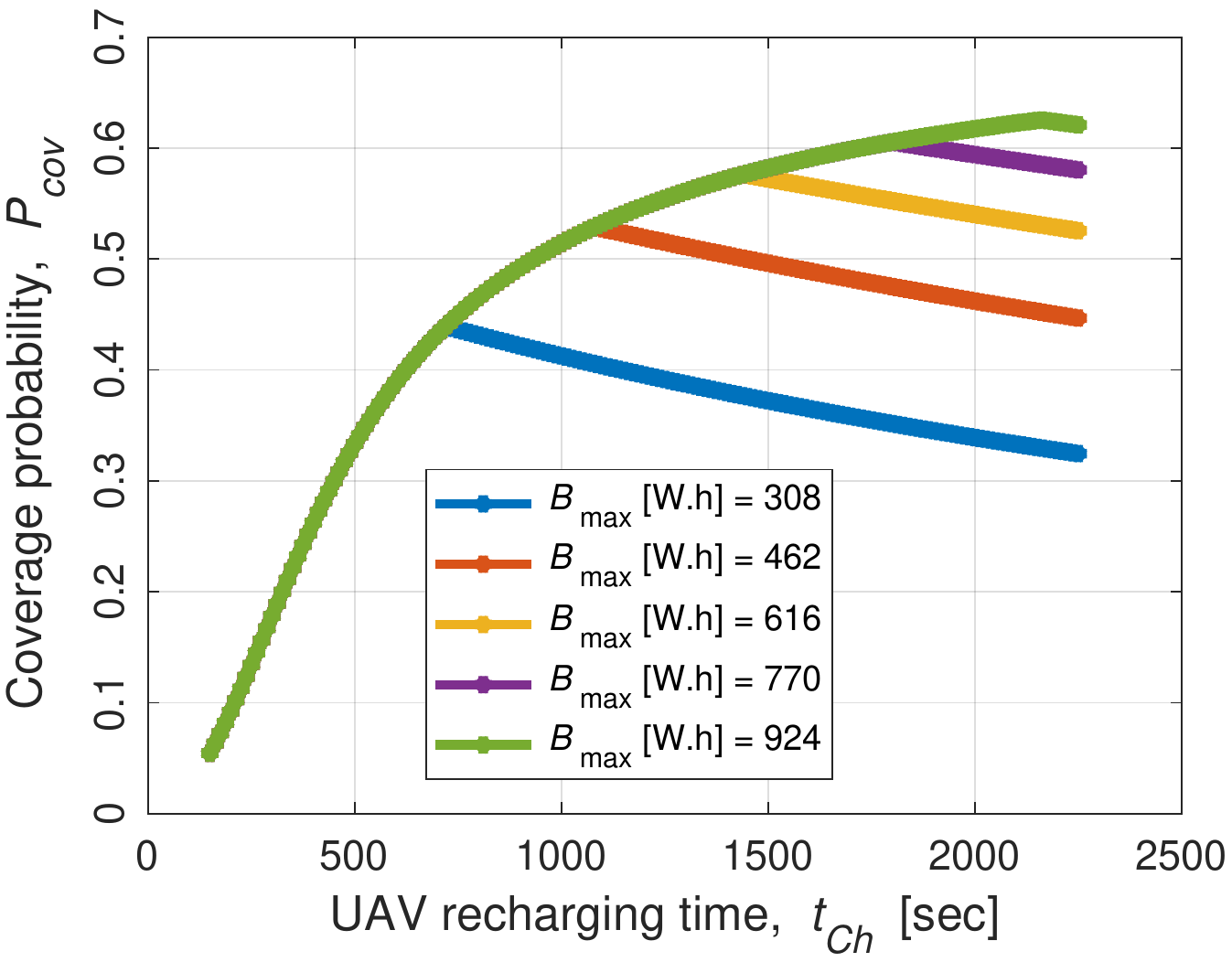}
    \vspace{0.8mm}
		\caption{}
	\end{subfigure}
	\hspace{-4mm}
	\begin{subfigure}{0.51\linewidth}
		\includegraphics[width=0.98\textwidth]{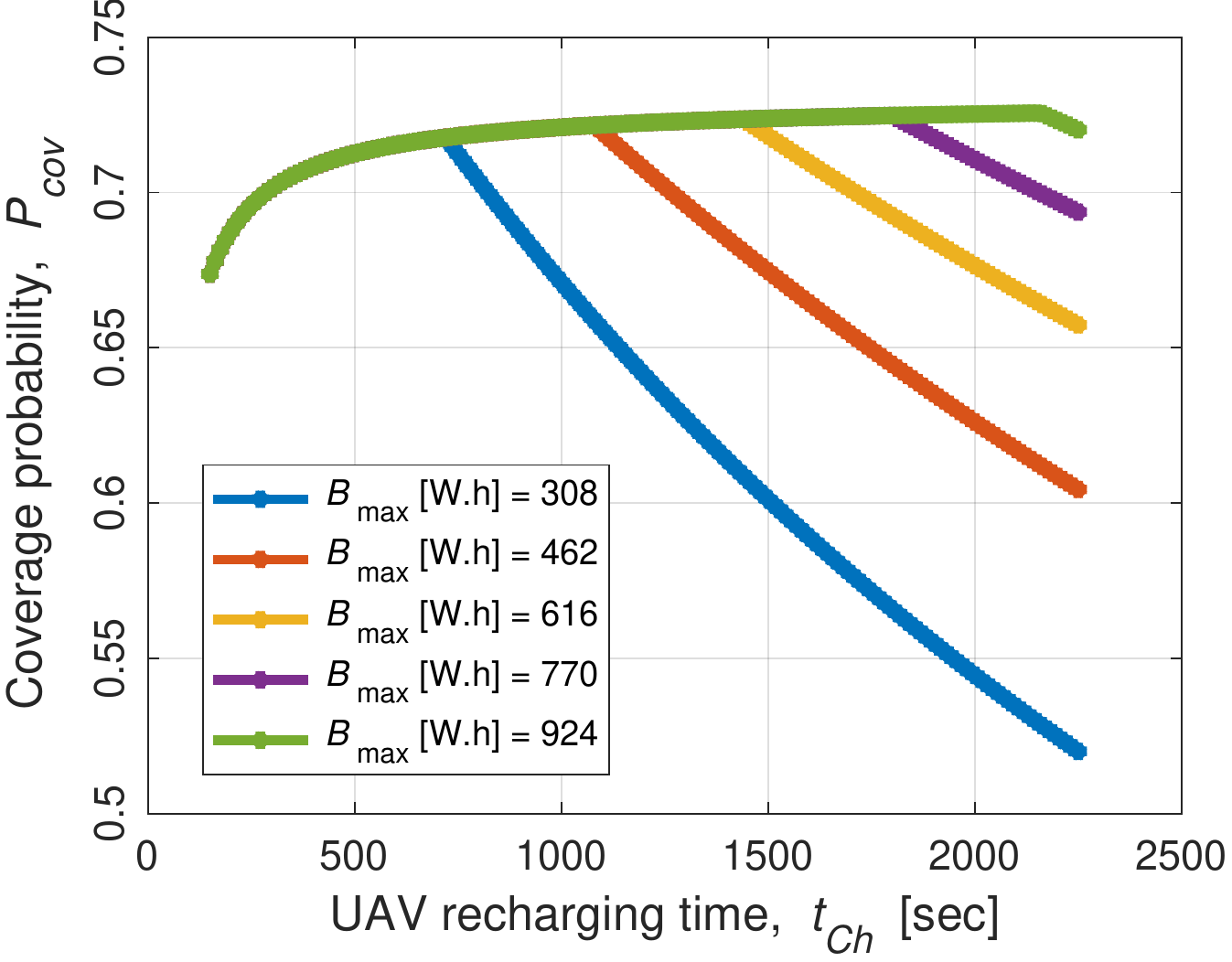}
    \vspace{1mm}
		\caption{}
	\end{subfigure}
    %\vspace{-1mm}
	\caption{Coverage probability, $P_{cov}$, vs. recharging time, $t_{Ch}$, for different $B_{\textrm{max}}$: (a) @$\lambda_{Ch}\!=\!10^{-9}\!$ [m\small$^{-2}$\normalsize]; (b) @$\lambda_{Ch}\!=\!10^{-6}\!$ [m\small$^{-2}$\normalsize].}  
	\label{fig:Pcov_vs_tch_forB}
	\vspace{-0.3mm}
\end{figure}

We first investigate the effect of varying recharging time on the coverage probability for an increasing recharging station density. For this analysis, using (\ref{eq:battery_charging})~for the considered $B_{\textrm{max}}$ and $\xi_{Ch}$ values, we know that $t_{Ch}^{sat}$ should be $3600$ seconds, i.e., $1$ hour, which is typical for commercial UAVs. Fig.~\ref{fig:Pcov_vs_lambda_for_tch}(a) depicts what happens when the UAV stays at the recharging station until its battery gets half-full gradually. %we observe that when the battery is half charged, the UAV performs relatively similar. 
Here, we observe that, depending on the recharging level, certain discrepancies arise, which are especially evident for lower densities of recharging stations. For example, in the case of $0.001$ stations per km$^{2}$, $P_{cov}$ gets $2.7$ times better when $t_{Ch}$ is increased from $600$ to $1800$ seconds. This effect is less apparent for higher $\lambda_{Ch}$ because finding a recharging station becomes more likely for the UAV; hence, it may not need to depend heavily on the charge in its battery due to the increased recharging possibility. Numerically speaking, $P_{cov}$ increases only up to $1.3$ times for the same increment in $t_{Ch}$ when $\lambda_{Ch}$ is ten times higher, i.e., $0.01$ per km$^{2}$. When we look at the fully charged case shown in Fig.{\ref{fig:Pcov_vs_lambda_for_tch}}(b), which can be considered ideal in theory, we see that the performance gap increases for lower-density values. Unsurprisingly, a fully-charged battery can help the UAV achieve better coverage, especially at a low $\lambda_{Ch}$, compared to those of partially-charged cases. Finally, in Fig.{\ref{fig:Pcov_vs_lambda_for_tch}}(c), we analyze the case when the battery overflowed, i.e., when the UAV continues to stay at the recharging station after its battery gets fully charged. From (\ref{eq:1}), we know that increasing $t_{Ch}$ beyond saturation is unsuggested since the service probability, $P_e$, and hence $P_{cov}$, is inversely proportional to it. However, Fig.{\ref{fig:Pcov_vs_lambda_for_tch}}(c) reveals that this might not be the case always. As seen, when the battery is charged for $4350$ seconds ($>\!t_{Ch}^{sat}$), it is still possible to obtain higher $P_{cov}$ compared to the partial charge cases, depending on $\lambda_{Ch}$. For example, $t_{Ch}$ of $600$ seconds can only outperform the overflowed case when $\lambda_{Ch}$ is $7.81\times10^{-6}$m$^{-2}$ or higher. Similar comments also hold for other $t_{Ch}$ values (except for $t_{Ch}^{sat}$), as can be seen from the dashed red lines perpendicular to the \textit{x}-axis, where the $\lambda_{Ch}$ threshold significantly reduces for increasing charge level towards the full battery, e.g., $2\times10^{-9}$m$^{-2}$ for $t_{Ch}$ of $2850$ seconds. The take-home message from this analysis is that a relatively empty battery might be worse than overstaying at the recharging station, e.g., for maintenance purposes, if the application mandates a certain level of $\lambda_{Ch}$.\\ 
%\begin{figure}[!h]
%	\centering
%	\includegraphics[width=0.67\columnwidth]{figs/4_cropped.pdf}
%	\caption{Coverage probability, $P_{cov}$, as a function of recharging time, $t_{Ch}$, for different battery sizes, $B_{\textrm{max}}$.}
%	\label{fig:h_N_D}
%\end{figure}
\indent Next, we extend our discussion to the impact of using batteries of different sizes on the coverage probability for varying charging times. As can be seen from Fig.~\ref{fig:Pcov_vs_tch_forB}(a), in the case of a partial charge, all batteries show the same performance until $t_{Ch}=1440$ seconds, i.e., when the battery of size $308$Wh gets full. That is because all batteries are charged to the same level at that $t_{Ch}$ value regardless of their total size, $B_{\textrm{max}}$. The same phenomenon is also observed for the remaining batteries until the one with the smallest size reaches its $t_{Ch}^{sat}$, where the respective $P_{cov}$ starts its dramatic decay due to overflow. Fig.~\ref{fig:Pcov_vs_tch_forB}(a) and Fig.~\ref{fig:Pcov_vs_tch_forB}(b) are for different densities of stations; $10^{-9}$ and $10^{-6}$ [m\small$^{-2}$\normalsize], respectively. Although $t_{Ch}^{sat}$ of each battery remains the same, the achievable $P_{cov}$ changes dramatically due to the clockwise rotation of the behavior. %revealing that the system designer 

\begin{figure}[t!]
    \centering
    \includegraphics[width=0.9\linewidth]{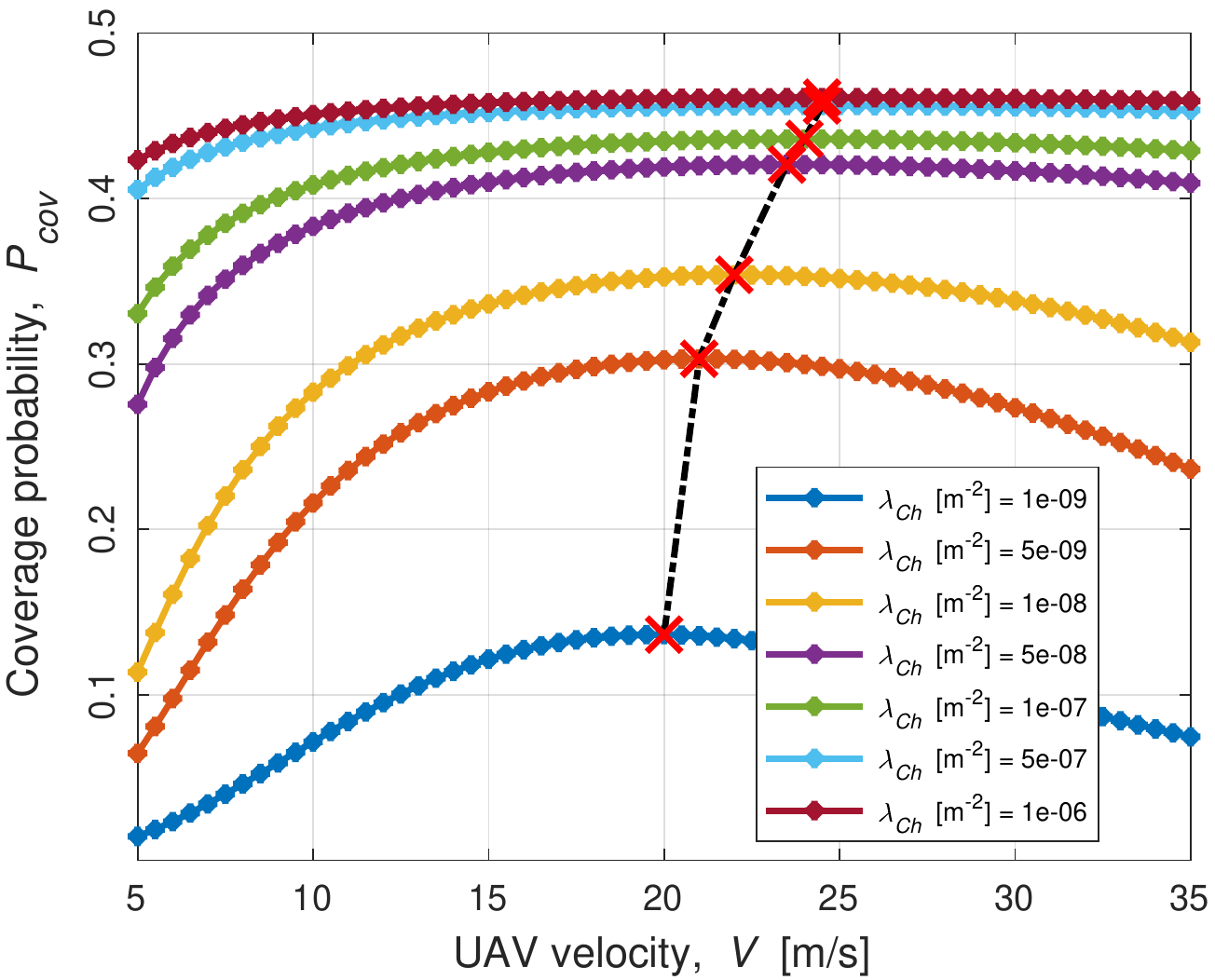}
    %\vspace{-1mm}
    \caption{Coverage probability, $P_{cov}$, as a function of UAV velocity, $V$, for varying $\lambda_{Ch}$ values.}
    \label{fig:velocity_eff}
    %\vspace{-6mm}
\end{figure}

We lastly focus on understanding how the UAV velocity affects the coverage performance depending on the recharging station density. Before explaining the results, we should note that Fig.~\ref{fig:velocity_eff} is produced for a $B_{\textrm{max}}$ of $192.5$Wh, i.e., the quarter of what Table~\ref{tab:parameters} mentions, which is only charged until its half at the same rate. As seen, the impact of $V$ is negligible for high $\lambda_{Ch}$ due to the same reason~explained earlier in this section. For low $\lambda_{Ch}$, on the other hand, the UAV should speed up to reach the nearest station faster, although that means an increased $P_J$. Due to the $P_J$ component, however, the increment in $V$ is only beneficial until a certain point, as shown with the black dashed line marking the optimal velocity values for each $\lambda_{Ch}$. Using a bigger battery and charging it to the ideal level, i.e., $100\%$ -no overflow, will diminish this behavior, forcing higher speeds only for low $\lambda_{Ch}$ values.

%\newpage
The results above suggest that for a given performance target, e.g., maximum $P_{cov}$ with minimum $\lambda_{Ch}$, shorter $t_{Ch}$, or a fixed $B_{\textrm{max}}$, the relevant design parameters can be tweaked as required. Considering that most of these parameters affect each other, e.g., the EIRP limit on $P_T$ alters not only $P_{\mathrm{LoS}}$ (due to the dictated $\theta_B$) but also the size of the event area and hence the number of sensors that can be powered (not within the scope of this study) and even their reporting frequencies, the network requires a holistic design approach as optimizing the trade-offs for the best performance achievable. We believe this paper provides a practice-based showcase on this, providing guidance for future efforts. 

%The results above suggest that optimal selection of relevant design parameters is of utmost importance to meet the requirements (e.g., minimum $\lambda_{Ch}$, maximum $P_{cov}$, shorter $t_{Ch}$) set by applications. Considering the fact that most of the parameters affect each other, e.g., the EIRP limit on $P_T$  
%certain reporting frequency, the number of sensors covered, highest throughput)

%\textcolor{red}{\lipsum[4-6]}

%\textcolor{blue}{As can be observed from Fig. 1, the charging time has a significant impact on the value of the density of charging stations required to achieve a specific value of coverage probability. For instance, the value of Pcov achieved with 1 charging station/km2 with charging time of 40 minutes, can be achieved with 100 times less density of charging stations if we can provide more efficient charging stations that have a charging time of 5 minutes. Similar comments also hold for the influence of the battery size Bmax on the the value of the density of charging stations required to achieve a given value of Pcov.}

\section{Conclusions}
%\textcolor{red}{\lipsum[5]}
\label{sec:conclusions}

This study investigates the factors affecting the coverage performance of sensors energized by UAVs with directional antennae, constituting an energy-neutral IoT network together with the multi-source EH recharging stations replenishing UAVs batteries. With this goal, We first derived the service probability as a function of UAV power consumption/velocity and battery size, recharging time and station density, and WPT duration. That was then joined with distance-conditioned coverage probability, incorporating the effects of directivity, the LoS/NLoS connectivity, and the non-linear EH model.  
%First, the maximum sensing range of the w-pDs was derived as a function of UAV altitude, output power, and directivity. Then, by using the CPP theorem, an efficient deployment strategy avoiding the overlaps and oversteps, and thus providing a lower-bound for the required number of w-pDs was proposed. 
%In this letter, we derived the coverage probability for a UAV-assisted cellular network as a function of the battery size, the density of the UAV charging stations, and the charging time. Using numerical results, we showed the high impact of the aforementioned system parameters on the system performance. One of the main drawn insights from this letter is the trade-off between deploying high density of low quality charging stations (high charging time) and deploying low density of high quality charging stations (low charging time). Our results showed that we could achieve similar coverage probability with lower density of charging stations if we can reduce the charging time.
The analyses revealed the design considerations for the best coverage with regard to FCC regulations, realistic rectenna and UAV operation, and minimum power requirements of sensors. 
Future works will focus on %determining the number of sensors that can be powered depending on the size of event area and the deployment strategy of sensors, i.e., random or well-planned/grid-type. We will also investigate how to 
%deriving an optimization problem to 
calculating the number of sensors that can be powered by UAVs. We will also try maximizing the communication throughput of sensors in the envisioned setting %, i.e., UAV-powered energy-neutral networks, 
by optimizing the parameters/trade-offs that have been discussed in this study.
%in such energy-neutral networks powered by UAVs by optimizing the design parameters that have been discussed in this study.

%\vspace*{\fill}

\section{Acknowledgement}
This work has been supported by the PETRAS National Centre of Excellence for IoT Systems Cybersecurity, funded by the UK EPSRC under grant number EP/S035362/1.
%This work is supported by the UK EPSRC under EP/S035362/1. Simulation data will be made available online once this paper is accepted.

\bibliographystyle{IEEEtran}
\bibliography{References}

% Generated by IEEEtran.bst, version: 1.14 (2015/08/26)
\begin{thebibliography}{10}
\providecommand{\url}[1]{#1}
\csname url@samestyle\endcsname
\providecommand{\newblock}{\relax}
\providecommand{\bibinfo}[2]{#2}
\providecommand{\BIBentrySTDinterwordspacing}{\spaceskip=0pt\relax}
\providecommand{\BIBentryALTinterwordstretchfactor}{4}
\providecommand{\BIBentryALTinterwordspacing}{\spaceskip=\fontdimen2\font plus
\BIBentryALTinterwordstretchfactor\fontdimen3\font minus
  \fontdimen4\font\relax}
\providecommand{\BIBforeignlanguage}[2]{{%
\expandafter\ifx\csname l@#1\endcsname\relax
\typeout{** WARNING: IEEEtran.bst: No hyphenation pattern has been}%
\typeout{** loaded for the language `#1'. Using the pattern for}%
\typeout{** the default language instead.}%
\else
\language=\csname l@#1\endcsname
\fi
#2}}
\providecommand{\BIBdecl}{\relax}
\BIBdecl

\bibitem{baltaci2021survey}
A.~Baltaci \emph{et~al.}, ``{A Survey of Wireless Networks for Future Aerial
  Communications (FACOM)},'' \emph{IEEE Communications Surveys \& Tutorials},
  vol.~23, no.~4, pp. 2833--2884, 2021.

\bibitem{cetinkaya2020efficient}
O.~Cetinkaya and G.~V. Merrett, ``{Efficient Deployment of UAV-powered Sensors
  for Optimal Coverage and Connectivity},'' in \emph{IEEE
  Wireless~Communications and$\,$Networking$\,$Conference$\,$(WCNC)}, 2020, pp.
  1--6.

\bibitem{yuan2021joint}
X.~Yuan \emph{et~al.}, ``Joint {D}esign of {UAV} {T}rajectory and {D}irectional
  {A}ntenna {O}rientation in {UAV}-enabled {W}ireless {P}ower {T}ransfer
  {N}etworks,'' \emph{IEEE Journal on Selected Areas in Communications},
  vol.~39, no.~10, pp. 3081--3096, 2021.

\bibitem{long2018energy}
T.~Long \emph{et~al.}, ``{Energy Neutral Internet of Drones},'' \emph{IEEE
  Communications Magazine}, vol.~56, no.~1, pp. 22--28, 2018.

\bibitem{HEH}
O.~B. Akan \emph{et~al.}, ``Internet of {H}ybrid {E}nergy {H}arvesting
  {T}hings,'' \emph{IEEE Internet of Things Journal}, vol.~5, no.~2, pp.
  736--746, 2018.

\bibitem{alzenad20173}
M.~Alzenad \emph{et~al.}, ``3-{D} {P}lacement of an {U}nmanned {A}erial
  {V}ehicle {B}ase {S}tation ({UAV-BS}) for {E}nergy-efficient {M}aximal
  {C}overage,'' \emph{IEEE Wireless Communications Letters}, vol.~6, no.~4, pp.
  434--437, 2017.

\bibitem{xie2021uav}
L.~Xie \emph{et~al.}, ``{UAV}-enabled {W}ireless {P}ower {T}ransfer: {A}
  {T}utorial {O}verview,'' \emph{IEEE Transactions on Green Communications and
  Networking}, vol.~5, no.~4, pp. 2042--2064, 2021.

\bibitem{ye2020optimization}
H.-T. Ye \emph{et~al.}, ``{O}ptimization for {W}ireless-powered {I}o{T}
  {N}etworks enabled by an {E}nergy-limited {UAV} {U}nder {P}ractical {E}nergy
  {C}onsumption {M}odel,'' \emph{IEEE Wireless Communications Letters},
  vol.~10, no.~3, pp. 567--571, 2020.

\bibitem{qin2020performance}
Y.~Qin \emph{et~al.}, ``{Performance Evaluation of UAV-enabled Cellular
  Networks with Battery-limited Drones},'' \emph{IEEE Communications Letters},
  vol.~24, no.~12, pp. 2664--2668, 2020.

\bibitem{zhang2022energy}
L.~Zhang \emph{et~al.}, ``{Energy-Efficient Trajectory Optimization for
  UAV-Assisted IoT Networks},'' \emph{IEEE Transactions on Mobile Computing},
  vol.~21, no.~12, pp. 4323--4337, 2022.

\bibitem{hassija2020adistributed}
V.~Hassija \emph{et~al.}, ``{A Distributed Framework for Energy Trading Between
  UAVs and Charging Stations for Critical Applications},'' \emph{IEEE
  Transactions on Vehicular Technology}, vol.~69, no.~5, pp. 5391--5402, 2020.

\bibitem{chu2022joint}
N.~H. Chu \emph{et~al.}, ``{Joint Speed Control and Energy Replenishment
  Optimization for UAV-assisted IoT Data Collection with Deep Reinforcement
  Transfer Learning},'' \emph{IEEE Internet of Things Journal}, pp. 1--1, 2022.

\bibitem{wei2022uav}
Z.~Wei \emph{et~al.}, ``{UAV}-assisted {D}ata {C}ollection for {I}nternet of
  {T}hings: {A} {S}urvey,'' \emph{IEEE Internet of Things Journal}, vol.~9,
  no.~17, pp. 15\,460--15\,483, 2022.

\bibitem{balanis2015antenna}
C.~A. Balanis, \emph{{A}ntenna {T}heory: {A}nalysis and {D}esign}.\hskip 1em
  plus 0.5em minus 0.4em\relax John wiley \& sons, 2015.

\bibitem{2010}
``Federal {{Communications Commission CFR}}, {{Title}} 47, {{Volume}} 1,
  {{Part}} 15,'' https://www.govinfo.gov/app/details/CFR-2010-title47-vol1,
  2010.

\bibitem{cetinkaya2017}
O.~Cetinkaya and O.~B. Akan, ``Electric-{{Field Energy Harvesting From Lighting
  Elements}} for {{Battery}}-{{Less Internet}} of {{Things}},'' \emph{IEEE
  Access}, vol.~5, pp. 7423--7434, 2017.

\bibitem{zeng2019energy}
Y.~Zeng \emph{et~al.}, ``{Energy Minimization for Wireless Communication with
  Rotary-wing UAV},'' \emph{IEEE Transactions on Wireless Communications},
  vol.~18, no.~4, pp. 2329--2345, 2019.

\bibitem{wagih2020high}
M.~Wagih \emph{et~al.}, ``{H}igh-efficiency {s}ub-1 {GH}z {F}lexible {C}ompact
  {R}ectenna based on {P}arametric {A}ntenna-rectifier {C}o-design,'' in
  \emph{2020 IEEE/MTT-S International Microwave Symposium (IMS)}, 2020, pp.
  1066--1069.

\bibitem{alevizos2018nonlinear}
P.~N. Alevizos \emph{et~al.}, ``{N}onlinear {E}nergy {H}arvesting {M}odels in
  {W}ireless {I}nformation and {P}ower {T}ransfer,'' in \emph{2018 IEEE 19th
  International Workshop on Signal Processing Advances in Wireless
  Communications (SPAWC)}.\hskip 1em plus 0.5em minus 0.4em\relax IEEE, 2018,
  pp. 1--5.

\bibitem{li2019joint}
J.~Li \emph{et~al.}, ``{J}oint {O}ptimization on {T}rajectory, {A}ltitude,
  {V}elocity, and {L}ink {S}cheduling for {M}inimum {M}ission {T}ime in
  {UAV}-aided {D}ata {C}ollection,'' \emph{IEEE Internet of Things Journal},
  vol.~7, no.~2, pp. 1464--1475, 2019.

\bibitem{liao2021hotspot}
Z.~Liao \emph{et~al.}, ``{HOTSPOT}: {A} {UAV}-assisted {D}ynamic
  {M}obility-aware {O}ffloading for {M}obile-edge {C}omputing in 3-{D}
  {S}pace,'' \emph{IEEE Internet of Things Journal}, vol.~8, no.~13, pp.
  10\,940--10\,952, 2021.

\bibitem{almarhabi2021lora}
A.~Almarhabi \emph{et~al.}, ``Lo{R}a and {H}igh-altitude {P}latforms: {P}ath
  {L}oss, {L}ink {B}udget and {O}ptimum {A}ltitude,'' in \emph{2020 8th
  International Conference on Intelligent and Advanced Systems (ICIAS)}.\hskip
  1em plus 0.5em minus 0.4em\relax IEEE, 2021, pp. 1--6.

\end{thebibliography}

%\noindent \underline{Questions?}\\

%1. What should be the height of recharging stations? Do we have that in \cite{al2014optimal}? The reference article took UAV height as $60$m.

%\textcolor{blue}{Biz drone'larla sensorlere power transfer edip datalarini topluyoruz. Research question: minimum UAV kullanarak en az outage'i nasil saglariz? Ayni anda ground'da da maximum sensing region'i basaramaya calisiyoruz. Drone'lar enerjilerini multi-source energy harvest eden recharging station'lardan aliyor. Battery replenishment suresince power ya da data transfer servisi yok, outage sebebi de bu. Optimizasyon olmayacak, tamamen analitik. Drone'larin sonlu bir kaynaktan power alip transfer etmesi ve base station olarak calismasi ilgi cekiyor.} 

\end{document}